\documentclass[10pt,letterpaper]{article}

\usepackage{graphicx} 
\usepackage[T1]{fontenc}
\usepackage{amsmath,amssymb,amsthm,mathtools}
\usepackage{newtxtext,newtxmath}
\usepackage{bm}
\usepackage{microtype}

\usepackage{algorithm}
\usepackage{algpseudocode}
\usepackage{url}
\usepackage{hyperref}
\usepackage[numbers,sort&compress]{natbib}

\usepackage{booktabs}

\usepackage{authblk}

\usepackage[
  letterpaper,
  left=0.85in,
  right=0.85in,
  top=0.74in,
  bottom=0.82in,
  footskip=0.35in
]{geometry}

\usepackage[nameinlink, capitalize]{cleveref}

\newcommand{\sep}{\mathrm{sep}}

\theoremstyle{thmstyleone}%
\newtheorem{theorem}{Theorem}%

\newtheorem{definition}{Definition}
\newtheorem{proposition}{Proposition}

\begin{document}

\title{A reduction scheme for general-order Ising-like Hamiltonians in quantum heuristic solvers}

\author[1]{Chengsi Mao}
\author[2]{Pavel Mosharev}
\author[2]{Yao Wang\thanks{wangyao123@huawei.com}}
\author[2]{Man-Hong Yung\thanks{yung@iqasz.cn}}

\affil[1]{State Key Laboratory of Surface Physics, Department of Physics, and Center for Field Theory and Particle Physics, Fudan University, Shanghai, 200433, China}
\affil[2]{2012 Laboratories, Huawei Technologies Co., Ltd., Shenzhen, 518129, China}

\maketitle

\begin{abstract}
The Ising model is ubiquitous in various optimization problems but notoriously difficult to solve due to combinatorial explosion. In view of this, Hamiltonian reduction is a useful preprocessing technique for reducing the effective problem size before applying heuristic solvers. However, existing reduction techniques mainly target second-order Ising models, whereas many pseudo-Boolean formulations naturally contain higher-order interactions. In this work, we generalize the concept of non-separable groups to arbitrary-order Ising-like models and develop a Hamiltonian reduction framework that iteratively detects and merges constrained spin groups into single variables. We benchmark the reduction on synthetic hypergraphs and higher-order network datasets, and evaluate its integration with downstream order-reduction and solver workflows. Our results establish a foundation for Hamiltonian reduction in higher-order Ising-like optimization problems.
\end{abstract}

\section{Introduction}

The last few years have witnessed a rapid development of quantum and quantum-inspired computing in various hardware implementations and algorithmic frameworks. One of the most promising avenues is to encode optimization problems into Ising models, whose ground states correspond to the optimal solutions~\cite{lucas2014ising,mohseni2022ising}. This field has been the focus of much attention since many combinatorial problems, including the well-known Karp's 21 NP-complete problems~\cite{Karp1972}, can be formulated into Ising Hamiltonians~\cite{lucas2014ising}. Other examples include scientific and industrial applications such as large-scale integrated circuit design~\cite{barahona1988application} and drug design~\cite{sakaguchi2016boltzmann}.

In general, finding the ground state of an Ising Hamiltonian is NP-hard, so it is difficult to solve on classical computers due to combinatorial explosion~\cite{barahona1982computational,arora2009computational}. Quantum computers, though promising, remain limited in scale, hindering their application to large optimization instances. For instance, existing quantum annealers such as D-Wave Advantage2 contain only a few thousand physical qubits and have constrained connectivity~\cite{dwave_docs_advantage2_2025}. Gate-based quantum processors are also limited by qubit count, connectivity, and noise, which restricts their ability to solve large optimization instances directly.
In view of this, variable fixing and Hamiltonian reduction techniques have been proposed and studied~\cite{gueye2025}, two prominent examples being roof duality~\cite{boros2006preprocessing, boros2002pseudo} and FastHare (FH)~\cite{thai2022fasthare}. Roof duality aims to find a partial assignment to binary variables in quadratic unconstrained binary optimization (QUBO) problems, an equivalent form to the second-order Ising problems. In contrast, FastHare introduced the concept of the non-separable group (NG), defined as a subset of spins that have same fixed values in all optimal solutions. A group of non-separable spins can be merged into one, thus reducing the number of logical variables in the problem.

Common heuristic approaches for solving Ising problems include simulated annealing~\cite{kirkpatrick1983optimization,isakov2015optimised}, quantum annealing~\cite{kadowaki1998quantum,das2008colloquium,hauke2020perspectives}, dynamical system evolution such as simulated bifurcation (SB)~\cite{goto2019combinatorial, goto2021high}, coherent Ising machines~\cite{inagaki2016coherent,mcmahon2016fully,yamamoto2017coherent}, quantum adiabatic optimization~\cite{farhi2001quantum,albash2018adiabatic} and hybrid quantum-classical algorithms executed on universal gate-based devices~\cite{blekos2024, zhu2026combinatorial}. Most of these methods are designed to solve Ising models containing only pairwise interactions and local fields~\cite{mohseni2022ising}. However, important classes of optimization problems, such as satisfiability problems, map more seamlessly to Ising-like models with higher-order interactions~\cite{boros2002pseudo,biamonte2008nonperturbative,babbush2013resource}. A few methods for directly solving higher-order Ising models have also been studied recently, including a specialized version of SB~\cite{kanao2023}, higher-order Ising machines~\cite{bybee2023efficient, Prova2026}, adaptations of QAOA to hardware-compatible problem instances with localized higher-order terms~\cite{Pelofske2024short, Pelofske2024scaling}, and quantum-inspired annealing approaches based on gauge-symmetry-preserving formulations of higher-order binary optimization~\cite{Wang2025}.

Besides general-order solvers, another common approach for solving Ising-like problems is to first reduce the original problem to second order using order reduction (OR) techniques~\cite{boros2014quadratization,anthony2017quadratic,mandal2020compressed}, and then solve it with second-order solvers. Despite the use of the same word "reduction", order reduction aims to convert higher-order polynomial objective into an equivalent quadratic Ising or QUBO formulation by introducing auxiliary variables, while Hamiltonian reduction, in contrast, aims to decrease the number of logical variables in the original higher-order objective by identifying variables whose ground-state behavior is constrained.

In this landscape of heuristics and preprocessing methods for arbitrary-order Ising-like models, specialized Hamiltonian reduction techniques for higher-order interactions have not been explored previously, as both FastHare and roof duality directly apply only to second-order Ising problems. Thus the extension to higher orders is of both theoretical and practical interest.

Here, we propose the \textbf{General} \textbf{Ha}miltonian \textbf{Re}duction (GeneralHare, GH) scheme on the basis of generalizing non-separability theory. We prove the correctness of the reduction and evaluate its effectiveness on higher-order Ising-like models, including synthetic instances such as Erd{\H{o}}s-R{\'e}nyi and scale-free hypergraphs, as well as public higher-order network datasets such as contact and email networks. For benchmark tests on second-order Ising models, GeneralHare outperforms the commonly-used FastHare in terms of reduction ratio. We also examine how reduction interacts with downstream order-reduction workflows by comparing the variable overhead of direct quadratization, a version of our algorithm GH\_minimal followed by quadratization, quadratization followed by FastHare, and their combined pipeline. Additional solver-level experiments are reported in the Appendix~\cref{app:enhanced_solver}. 

Overall, these results indicate that GeneralHare is an effective preprocessing framework for reducing the logical size of higher-order Ising-like models while preserving ground-state correspondence through an explicit reconstruction map.

\section{Preliminaries}
This section introduces some preliminaries of this work, including the higher-order Ising-like model and several methods for solving the corresponding ground-state problem, as well as the concept of non-separable groups (NG).

\subsection{Higher-order Ising-like model}

\begin{figure}[h]
\centering
\includegraphics[width=0.3\textwidth]{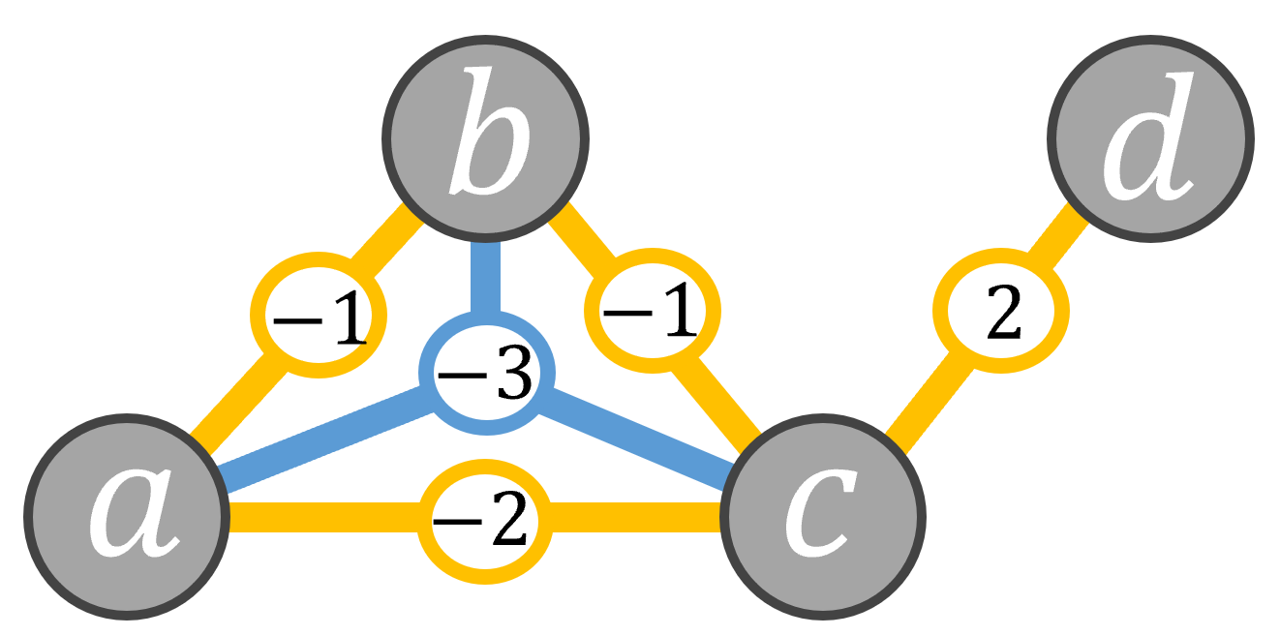}
\caption{Illustration of a hypergraph representing a higher-order Ising-like Hamiltonian. The numbers on the (hyper)edges denote the corresponding weights.}
\label{fig:GEJ}
\end{figure}

The Ising model describes a physical system of $n$ spins, with two-body interactions and an external field, where each spin $\sigma_i$ ($i=1,2,..,n$) takes the value of $\pm 1$. The Hamiltonian reads
\begin{equation}\label{eq:Ising}
	H(\boldsymbol{\sigma})
=
\sum_{i=1}^{n} h_i\sigma_i
+
\sum_{1\le i<j\le n} J_{ij}\sigma_i\sigma_j .
\end{equation}
When higher-order polynomial interactions are included, \cref{eq:Ising} can be generalized to the following form
\begin{equation}\label{eq:Ham}
H(\boldsymbol{\sigma})
=
\sum_{\emptyset \neq I \subseteq [n],\, |I|\le M}
J_I \prod_{i\in I}\sigma_i ,
\end{equation}
where $M$ is an integer, and $M \geq 2$. We shall call \cref{eq:Ham} the Hamiltonian of an order-$M$ Ising-like model. It can also be viewed as an order-$M$ hypergraph $G:=(V,E,J)$, where $V$, $E$ and $J$ denote the set of nodes, edges and weights, respectively. Every node $i \in V$ is associated with a discrete variable $\sigma_i \in \{-1,+1\}$ representing the spin on this node. The weight corresponding to hyperedge $I=\{i_1,...,i_m\} \, (I\in E)$ is $J_I$, and we assume no self-interactions present. \cref{fig:GEJ} shows a concrete example of a third-order hypergraph, which corresponds to the Hamiltonian $H=J_{abc}\sigma_a \sigma_b \sigma_c + J_{ab}\sigma_a \sigma_b + J_{bc}\sigma_b \sigma_c+ J_{ac}\sigma_a \sigma_c + J_{cd}\sigma_c \sigma_d +  J_a \sigma_a +J_b\sigma_b+J_c \sigma_c+J_d \sigma_d$. (For simplicity, we do not show linear terms in the graph.) For a subset of variables \(X\subseteq V\), an assignment \(\boldsymbol{\sigma}_X\in\{-1,+1\}^{|X|}\) is called a partial configuration on \(X\). If \(X=V\), it is called a full configuration. The configuration corresponding to the global minimum of $H(\boldsymbol{\sigma})$ is called ground state or optimal solution of the corresponding Ising-like problem. In numerical experiments with heuristic solvers, the lowest-energy configuration returned by the solver is referred to as the best-found solution.

\subsection{Hamiltonian reduction and non-separable groups}
In the course of solving Ising problems, researchers found that certain subsets of spins keep their relative configuration in all ground states of the given problem. The relative configuration here means that when going from one ground state to another, all spins in the subset can only be flipped together as a whole. This phenomenon was first reported and studied in~\cite{barahona1982morphology}, which focused on Ising models on square lattices, coining the term `solidary spins' for such subsets. The idea to replace a pair of spins that keep their relative configuration in all ground states by a single spin can also be found in~\cite{Zintchenko2015}.

Recently, a Hamiltonian reduction algorithm called FastHare (or FH for short) was proposed by applying similar ideas to arbitrary second-order Ising models~\cite{thai2022fasthare} and spin groups of arbitrary sizes. The authors define the `non-separable group' (NG) as a group of spins, that all have the same sign in all ground states of Hamiltonian $H$: $X$ is a non-separable group if and only if in all ground states of $H$, all spins in $X$ have value $+1$, or all spins in $X$ have value $-1$. If all spins in $X$ take the same value in some (but not all) ground states of Hamiltonian $H$, then $X$ is called a weakly non-separable group (weakly NG)~\cite{thai2022fasthare}. By recursively grouping NG (or weakly NG) into one spin, FastHare is able to reduce the original Hamiltonian into a smaller one, whose ground states correspond to the ground states of the original Hamiltonian via a simple map. In the original work~\cite{thai2022fasthare}, FastHare is shown to outperform roof duality implemented by D-Wave on various datasets. 

All methods mentioned above are designed for second-order Ising models, and not applicable to general-order problems. To bridge this gap, we extend the idea behind FastHare to higher-order Ising-like models. However, two main challenges need to be tackled in order to develop a generalized non-separability theory: inhomogeneity of Hamiltonian and the absence of correspondence to the min-cut problem. First, terms of different orders should be dealt with separately. While in second-order cases, the linear terms can be resolved by converting the problem into a Sherrington-Kirkpatrick (SK) graph~\cite{thai2022fasthare}, an analogue of the SK graph does not exist in higher orders, which adds to the difficulty of theoretical analysis. Secondly, the corresponding relation between searching ground states for Ising Hamiltonian and the weighted min-cut problem no longer holds in general-order cases, so a new approach for identifying non-separable groups needs to be developed. Developing a Hamiltonian reduction scheme for higher-order Ising-like models is therefore technically more challenging, but may provide substantial gains in problem-size reduction and solver efficiency.

\begin{figure}
    \centering
    \includegraphics[width=\textwidth]{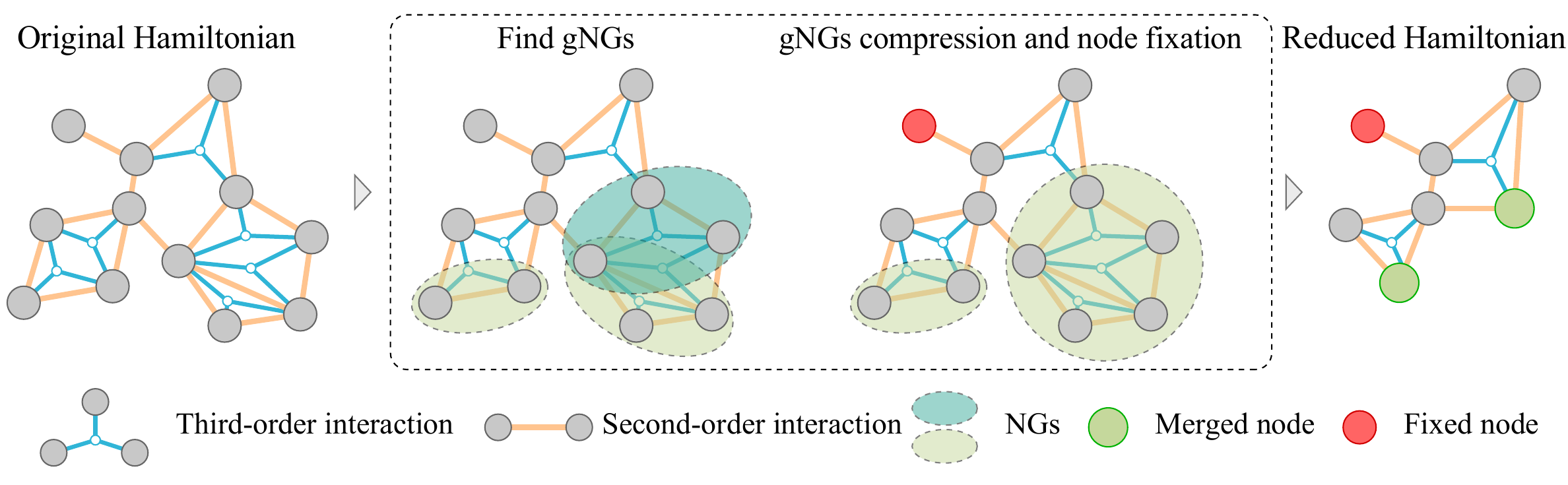}
    \caption{Diagrammatic illustration of the process of GeneralHare on higher-order Ising-like Hamiltonians. Here we set the highest order to be $M=3$ as an example.}
    \label{fig:schematic}
\end{figure}

\section{Results}
The main theoretical contribution of this work is that we generalize the non-separability theory of~\cite{thai2022fasthare} to higher-order Ising-like models. To this end, we introduce the `\textit{generalized non-separable group}'(gNG) to allow for arbitrary relative configurations inside a spin group to be merged, as well as deriving easily-computable criteria for identifying gNGs. See `Methods' for more details.
Based on the extended theories, we are able to propose the GeneralHare scheme, which iteratively detects and merges gNGs into single nodes until no more gNGs can be found. The diagrammatic illustration of the process of GeneralHare is shown in \cref{fig:schematic}.

To quantify the effectiveness of our Hamiltonian reduction scheme, we define the reduction ratio as $r=1-n_r/n$, where $n$ and $n_r$ are the node numbers for the original and reduced Hamiltonian, respectively. In the following, we test GeneralHare on generic higher-order Ising Hamiltonians as well as benchmarking on second-order cases by comparing with FastHare. Besides, we also test the combination of conservative demonstrator code G\_minimal with downstream quadratization and compare to more traditional FastHare-based approach. Additional experiments, including combination of GeneralHare with quantum-inspired heuristic solver, are present in Appendix~\cref{app:enhanced_solver}.

\begin{figure}[t]
\centering
\includegraphics[width=\textwidth]{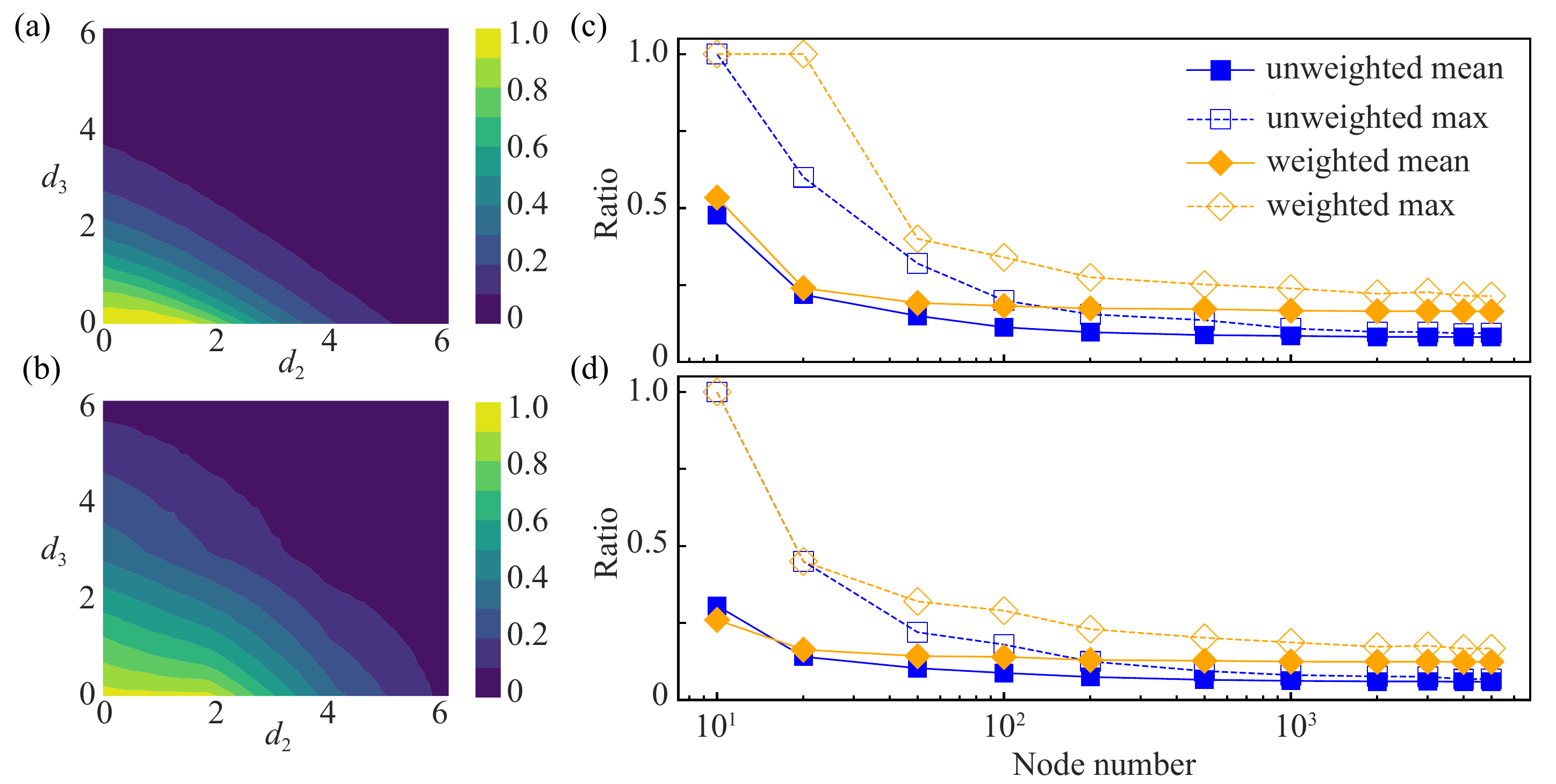}
\caption{\textbf{Reduction ratio on synthetic data }
     (The higher the better.)
    \textbf{(a)} Reduction ratio of third-order ER-like hypergraphs using GeneralHare and how it changes with the increase of second- and third-order node degree $d_2$, $d_3$, averaged over 100 random samples. The node number is $n=200$. For each hyperedge, $J$ is an integer chosen uniformly in $\{-4,-3,-2,-1,1,2,3,4\}$. 
    \textbf{(b)} Same as (a), but for third-order SF-like hypergraphs. 
    \textbf{(c)} Reduction ratio with respect to node number on third-order weighted (orange line) and unweighted (blue line) SF-like hypergraphs, with $d_2=4,d_3=2$. For weighted hypergraph, $J$ is an integer chosen uniformly in $\{-4,-3,-2,-1,1,2,3,4\}$ for all hyperedges, while for unweighted hypergraph, $J=1$ for all hyperedges. The corresponding hollow markers denote the highest reduction ratio in 100 random samples for each case. 
    \textbf{(d)} Same as (c), but for fourth-order SF-like hypergraphs, $d_2=4,d_3=2,d_4=1$. }
\label{fig:ratio}
\end{figure}

\subsection{Performance of GeneralHare}
\label{sec:performance}
To perform a comprehensive test on GeneralHare, we first experiment with synthetic datasets, which possess similar features but are parameterized and changeable. To do this, we generalize the (second-order) Erd{\H{o}}s-R{\'e}nyi~\cite{erdHos1960evolution} and scale-free~\cite{barabasi1999emergence,barabasi2002evolution,gay2005innovation,hanaki2010dynamics} graphs to hypergraphs, which are widely used in tests for Hamiltonian reduction. The algorithms we use for generating test instances are modified versions of those implemented in NetworkX library~\cite{hagberg2008exploring}. The details of generalization methods can be found in Appendix~\cref{app:data}.

\begin{table}[t]
\centering
\caption{\label{tab:real} \textbf{Features and reduction ratios of public higher-order network datasets.}}
\begin{tabular}{@{}lllllll@{}}
\toprule
Dataset name & Node number & Edge number & Highest order & Reduction ratio ($\%$) & Runtime (s)\\
\midrule
contact-primary-school & 242 & 12,799 & 5 & 2.1 & 8.7 \\
contact-high-school & 327 & 7,937 & 5 & 26.9 & 3.8 \\
DAWN & 2,558 & 143,523 & 16 & 67.4 & 1381.9\\
email-Enron & 143 & 1,542 & 18 & 9.1 & 0.3 \\
NDC-classes & 1,161 & 1,222 & 24 & 48.3 & 0.17 \\
email-Eu & 998 & 25,791 & 25 & 21.3 & 190.3\\
NDC-substances & 5,311 & 10,025 & 25 & 53.6 & 13.7 \\
\bottomrule
\end{tabular}
\end{table}

First, we investigate the influence of the graph density on GeneralHare. We change the average degree and obtain hypergraphs of different densities, and the benchmark results are shown in \cref{fig:ratio}(a,b). For both Erd{\H{o}}s-R{\'e}nyi (ER) and scale-free (SF) hypergraphs, the reduction ratio gradually decreases to zero as the average degree increases. We observe that sparser hypergraphs are easier to reduce by GeneralHare, which is consistent with the observation on the second-order graphs~\cite{thai2022fasthare}.

Second, to analyze the performance of GeneralHare on hypergraphs of different orders, we observe that the gradients of the reduction ratio with respect to interaction degrees differ across interaction orders. From \cref{fig:ratio} (a-b), we see that $\left| \frac{\partial r}{\partial d_3} \right| > \left|\frac{\partial r}{\partial d_2}\right|$, which implies that the higher the order $k$, the greater the influence $d_k$ has on reduction ratio. We also test with the third- and fourth-order SF-like hypergraphs, and the results are shown in \cref{fig:ratio} (c-d), respectively. The average reduction ratio of fourth-order hypergraphs is only about half of that of third-order ones, which demonstrates that higher-order hypergraphs display significantly lower reduction ratio than their lower-order counterparts. Overall, the higher-order interactions make the hypergraphs harder to reduce. 

Third, we can also see the influence of node number from these results. As shown in \cref{fig:ratio} (c), for a typical third-order SF-like hypergraph, the reduction ratio decreases from around $50\%$ to $20\%$ as the node number increases from 10 to 100, and remains almost constant as the node number continues to grow. 

Another factor that may affect the reduction ratio is the distribution of nodes and weights in the graph. Graphically, SF-like hypergraphs consist mainly of hubs (nodes that have many more connections than others) and branching structures, in contrast to ER-like hypergraphs in which the node degrees are more evenly distributed. By comparing results between \cref{fig:ratio} (a-b), we observe that unevenly distributed SF-like hypergraphs usually exhibit higher reduction ratios than evenly distributed ER-like hypergraphs. Apart from the hypergraph structure, we can also introduce hyperedge weights to make the weight distribution deviate from regular distribution. As shown in \cref{fig:ratio} (c-d), we find that GeneralHare is more effective on weighted hypergraphs, where the weights are unevenly distributed.

Furthermore, we test GeneralHare on public higher-order network datasets~\cite{benson2018}. These datasets include contact, email, drug-label, and drug-use simplicial-complex data, and therefore provide empirical higher-order topologies beyond synthetic random hypergraphs. They should be interpreted as topology benchmarks rather than complete industrial optimization objectives. In~\cref{tab:real}, we give the reduction ratio and runtime. GeneralHare achieves a reduction ratios from $2.1\%$ to $67.4\%$, exceeding $20\%$ on $5$ out of $7$ instances, which indicates that the method can be effective beyond synthetic instances.

\subsection{Comparison to FastHare in the second-order case}
\label{sec:fh_comparison}
As a benchmark, we compare the performance of GeneralHare with that of FastHare on second-order Hamiltonians in terms of reduction ratio and processing time. Since the non-separability theory behind GeneralHare can recover that of FastHare in the second-order case, they are expected to show similar reduction ratio and processing time on the same problems.

\begin{figure}[t]
    \centering
    \includegraphics[width=\linewidth]{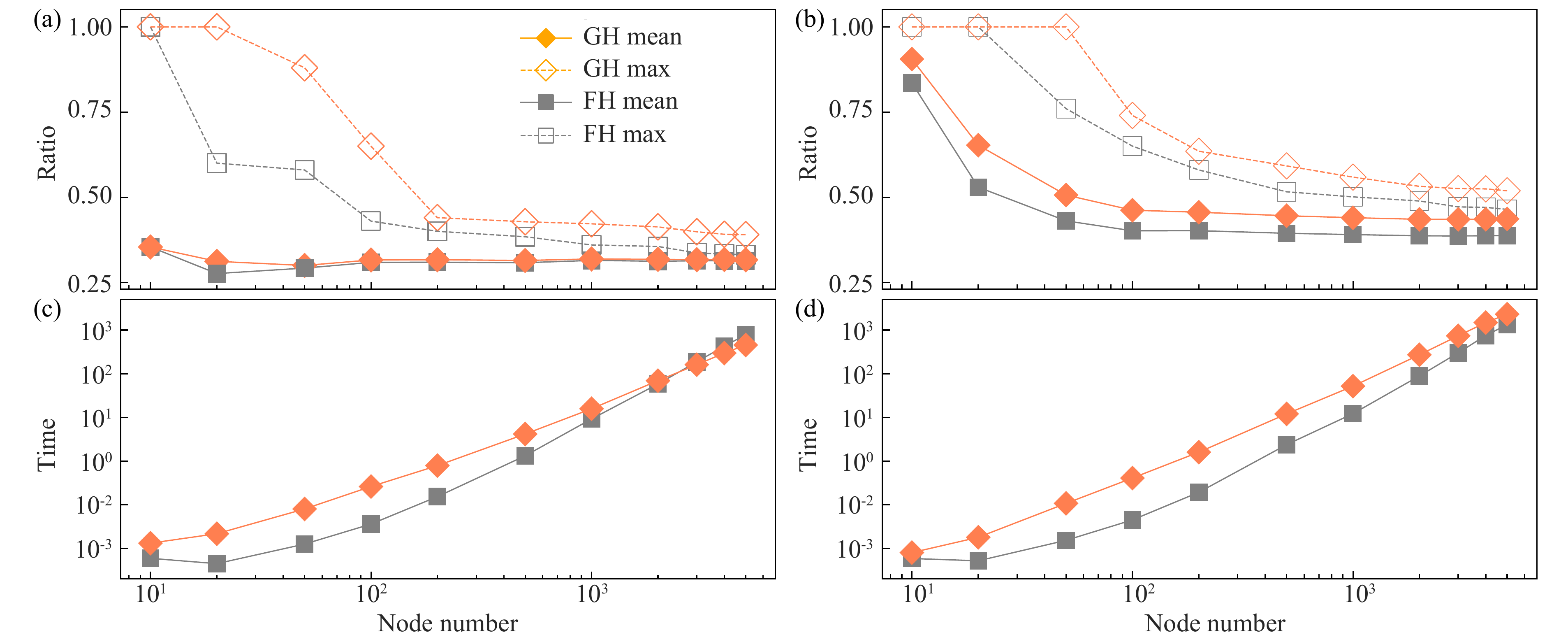}
    \caption{\textbf{Performance comparison of GeneralHare and FastHare on synthetic second-order Ising Hamiltonians.} \textbf{(a)} Reduction ratio of ER graphs; \textbf{(b)} same as (a), but for SF-like graphs. \textbf{(c)} Processing time for ER graphs; \textbf{(d)} same as (c), but for SF-like graphs. Solid markers denote the results averaged over 100 random instances at each problem size, and the corresponding hollow markers in (a) and (b) denote the highest reduction ratio obtained for these instances. `FH' means FastHare, and `GH' means GeneralHare.}
\label{fig:FHGH}
\end{figure}

Surprisingly, we find in numerical experiments that GeneralHare achieves higher reduction ratio than FastHare on both random ER and random SF-like graphs, as shown in~\cref{fig:FHGH}. This is because our scheme applies the criteria differently so that more gNGs are identified. Meanwhile, the additional criterion evaluation introduces a runtime overhead for GeneralHare, which is most visible on small graphs. To overcome this problem, several techniques for reducing the computational cost were developed, as introduced in the `Methods' section below, which effectively slow the increase of runtime for GeneralHare with respect to graph size. Specifically, we find a crossover in runtime comparison between GeneralHare and FastHare at the size of 5000 nodes as shown in \cref{fig:FHGH}(c).

\subsection{Combination of minimal GeneralHare demonstrator with order reduction}
\label{sec:pyqubo_or_combination}

Order reduction is one of the common ways to solve higher-order Ising formulations by converting higher-order model into quadratic form and applying quadratic solvers, including QUBO/Ising heuristics and quantum-annealing-oriented workflows. However, standard quadratization typically introduces auxiliary variables, so a higher-order model that is compact in its original form may become substantially larger after OR. This experiment therefore complements the direct reduction-ratio and runtime experiments by evaluating a downstream practical question: whether combining the quadratization with Hamiltonian Reduction can reduce the variable overhead of the resulting quadratic problem, and whether this benefit remains after an additional quadratic reduction step with FastHare. We use order reduction as implemented in PyQUBO~\cite{Zaman2022pyqubo} with default settings.

\begin{figure}[t]
    \centering
    \includegraphics[width=\linewidth]{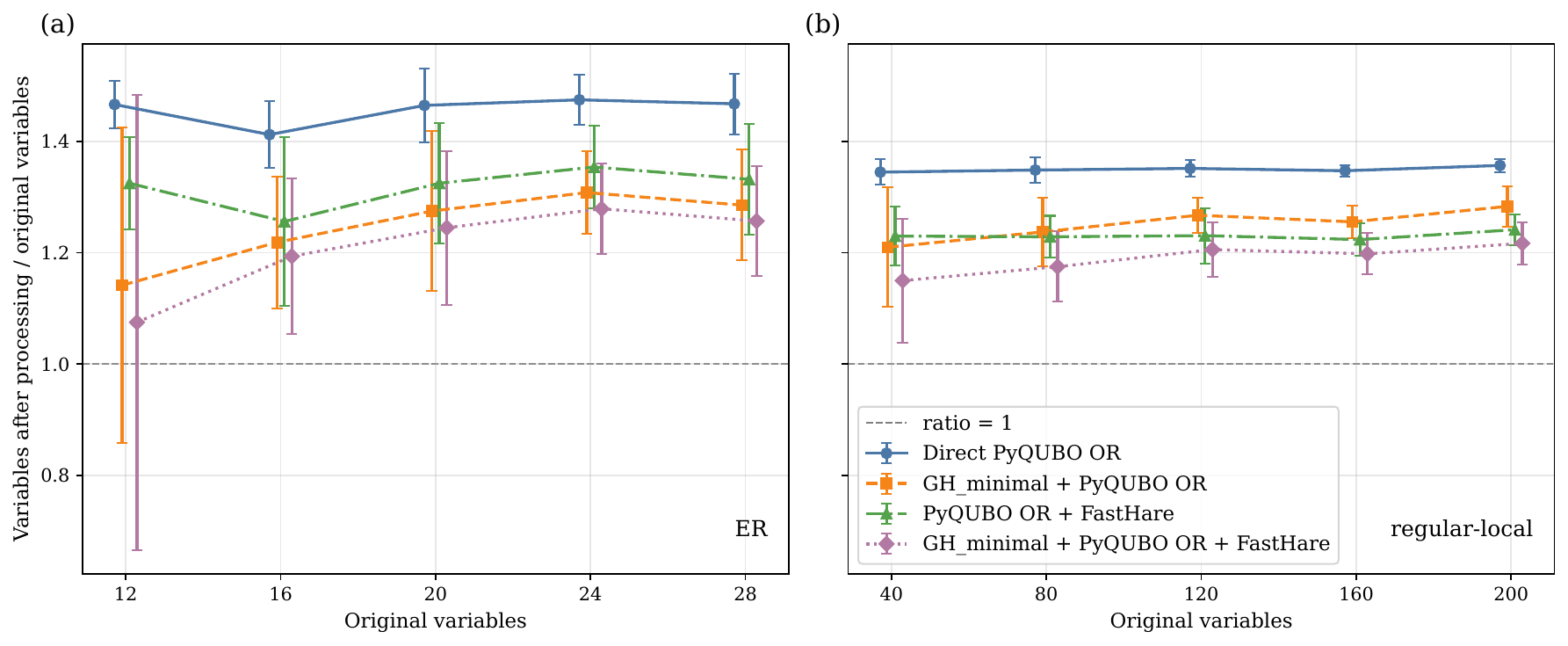}
    \caption{\textbf{Variable overhead of quadratization and reduction pipelines.} We use Python demonstrator implementation GH\_minimal, and report ratio of the number of variables after preprocessing to the original number of variables in the higher-order Ising Hamiltonian. Markers show averages over ten independently generated instances, and error bars show one standard deviation. We compare direct PyQUBO order reduction (OR), GH\_minimal followed by PyQUBO OR, PyQUBO OR followed by FastHare, and the combined GH\_minimal + PyQUBO OR + FastHare pipeline. The dashed horizontal line marks ratio 1, corresponding to no net variable overhead relative to the original higher-order problem size. \textbf{(a)} Ratio on ER-like hypergraphs; \textbf{(b)} Ratio on regular-local hypergraphs.}
\label{fig:OR}
\end{figure}

To evaluate the variable overhead introduced by quadratization and subsequent reduction, we generated synthetic higher-order Ising instances and compared four preprocessing pipelines: direct PyQUBO order reduction (OR), GH\_minimal followed by PyQUBO OR, PyQUBO OR followed by FastHare, and the combined GH\_minimal + PyQUBO OR + FastHare pipeline. All results in~\cref{fig:OR} were obtained with the released minimal Python demonstrator code, denoted GH\_minimal, which implements the conservative certified reduction core rather than the full original experimental implementation (see \textbf{Code availability}). 

For each system size, we generated ten independent random instances and report the mean and standard deviation of the ratio between the final variable count and the original number of higher-order variables. The ER-like instances used interaction orders \(2,3,4\) with target average degrees \(d_2=2.0\), \(d_3=1.0\), and \(d_4=0.5\). The regular-local instances used a random \(d_{\mathrm{reg}}=3\) regular pairwise backbone, with local higher-order target degrees \(d_3=0.6\) and \(d_4=0.4\). All generated hyperedges were assigned independent nonzero integer weights sampled uniformly from \(\{-4,-3,-2,-1,1,2,3,4\}\). GH\_minimal was run with candidate-size cutoff \(\xi=2\) and node fixation enabled. FastHare~\cite{thai2022fasthare} was applied with threshold parameter \(\alpha=1.0\). PyQUBO was used through the project-specific \texttt{compile\_hypergraph} interface with default penalty strength.

As we can see in~\cref{fig:OR}, application of GeneralHare even in the form of minimal conservative implementation GH\_minimal before order reduction results in smaller final variable overhead on the random ER-like graphs, compared to the standard way of applying FastHare after the quadratization. On both ER-like and regular-local graphs we can see that the best effect is achieved when the two Hamiltonian reduction schemes GH\_minimal and FastHare are applied together, before and after quadratization.

\section{Methods}

In this section, we will give a detailed statement on the generalized non-separability theory as well as a description on the GeneralHare scheme for arbitrary-order Ising-like models. The scheme features three ideas: first, we introduce the `\textit{generalized non-separable group}' to allow for arbitrary relative configurations inside a spin group to be merged; secondly, we derive lower bounds on the non-separability index for interactions of all orders, which sum up to the criteria we use; finally, we introduce a node-fixation stage, which fixes spins whose ground-state value can be certified locally. Besides, techniques including configuration pruning and simultaneous merging of weakly non-separable groups are developed to reduce the practical computational cost.

\subsection{Generalized non-separable groups}

\subsubsection{Non-separability theory}
Though originally defined on second-order Ising models, the definition of (weakly) NG applies to Ising-like models of general order.
As a concrete example, consider the Ising-like Hamiltonian in \cref{fig:GEJ}. If all linear terms are zero ($J_a=J_b=J_c=J_d=0$), then $X=\{a,b,c\}$ is an NG, with $\sigma_a=\sigma_b=\sigma_c=1$ in the ground state.
For completeness, here we first restate the non-separability conditions, some auxiliary concepts and results in the context of higher-order Ising-like models.

We can uniquely determine a configuration $\mathbf{\sigma}$ by a subset $S$ of $V$, i.e., for all $i\in S$, $\sigma_i=-1$, while for all $i \in V\backslash S$, $\sigma_i=1$.
Denote the number of common indices between hyperedge $I$ and $S$ by
\begin{equation}
	n_I(S):=|I\cap S|,
\end{equation}
and the sign of each term in $H(S)$ is defined by the parity of $n_I(S)$ of the corresponding hyperedge. By introducing notations $n_I(S):\text{EV}$ and $n_I(S):\text{OD}$ to denote that $n_I(S)$ is an even (odd) number, we can rewrite 
\cref{eq:Ham} for a spin configuration $S$ in the following equivalent forms
\begin{equation}\label{eq:energy}
	H(S) =\sum_{n_I(S):\text{EV}} J_I - \sum_{n_I(S):\text{OD}} J_I = \sum_{\text{all edges}} J_I - 2\sum_{n_I(S):\text{OD}} J_I.
\end{equation}

\begin{definition}[Separation]\label{def:sep}
	Consider a configuration $S$ and a subset $X \subseteq V$, we say $S$ separates $X$, denoted by $S \ominus X$, iff
	\begin{equation} 
		X \cap S \notin \{\emptyset, X\}.
	\end{equation}
	We also denote by $\sep(X)=\{S\subseteq V: S\ominus X\}$ the collection of all subsets that separate $X$.
\end{definition}

Then we can define the general non-separability index of $X$ as the difference between the minimum energy of the configurations in $\sep(X)$ and that of those outside $\sep(X)$, which applies for arbitrary-order Ising-like models, while in the second-order case, it recovers the original definition~\cite{thai2022fasthare}.
\begin{definition}[Non-separability index]\label{def:nu}
	Given a Hamiltonian $H$ and a subset $X \subseteq V$, the non-separability index of $X$ is defined as 
	\begin{equation}\label{eq:nu_def}
		\nu_H(X)=\frac{1}{2}\left[\min_{S \in \sep(X)} H(S) - \min_{S' \notin \sep(X)} H(S')\right].
	\end{equation}
\end{definition}
$\nu_H(X)$ is essentially (half of) the energy gap between the lowest energy among the configurations where not all spins in $X$ take the same value ($\min_{S \in \sep(X)} H(S)$), and its counterpart where all spins in $X$ take the same value ($\min_{S \notin \sep(X)} H(S)$).
The non-separability index plays a pivotal role since it is a direct indicator of (weakly) non-separable groups. We can give the conditions for non-separability as follows.
\begin{theorem}[Non-separability conditions]\label{thm:NG}
Given a group of spins $X \subseteq V$, 
\begin{itemize}
    \item $X$ is a non-separable group iff $\nu(X)>0$.
    \item $X$ is a weakly non-separable group iff $\nu(X)=0$.
\end{itemize}
\end{theorem}
This can be proved since $X$ being a non-separable group is equivalent to the statement that none of the ground state configurations can appear in $\sep(X)$, which further implies that 
\begin{equation}
	\min_{S \in \sep(X)} H(S) - \min_{S' \notin \sep(X)} H(S')=2\nu(X)>0.
\end{equation} 
Conversely, if $\nu(X)>0$, none of the ground states can appear in $\sep(X)$. In this way, we have proved the first statement, and the second statement follows with a similar argument.

From the definition of the non-separable groups, we can find that they have the following properties.
\begin{itemize}
	\item \emph{Hereditary}. If $X$ is a (weakly) non-separable group of $V$, then any subset $S \subseteq X$ is also a (weakly) non-separable group.
	\item \emph{Closure under union}. If $X$, $Y$ are non-separable groups, and $X\cap Y \neq \emptyset$, then $X \cup Y$ is also non-separable. If only one of $X$ or $Y$ is non-separable, the other is weakly non-separable, and $X\cap Y \neq \emptyset$, then $X \cup Y$ is weakly non-separable.
\end{itemize}

\subsubsection{Criteria for non-separability} 
While Theorem~\ref{thm:NG} gives the necessary and sufficient condition for non-separable groups, it cannot be used directly in the Hamiltonian reduction algorithm because evaluating $\nu_H(X)$ exactly requires a minimization over exponentially many spin configurations. Indeed, calculating \cref{eq:nu_def} exactly requires considering all spin configurations, which is similar to solving the whole problem with a brute force algorithm. For practical implementation, we resort to sufficient conditions that are easily computable. In what follows, we shall derive a lower bound $\hat{\nu}_H(X)$ on \cref{eq:nu_def}, which can be computed within time independent of the total node number $n$. Thus if $\hat{\nu}_H(X)>0$, we must have $\nu_H(X)>0$.

\begin{figure}[tbp]
	\centering
	\includegraphics[width=0.4\textwidth]{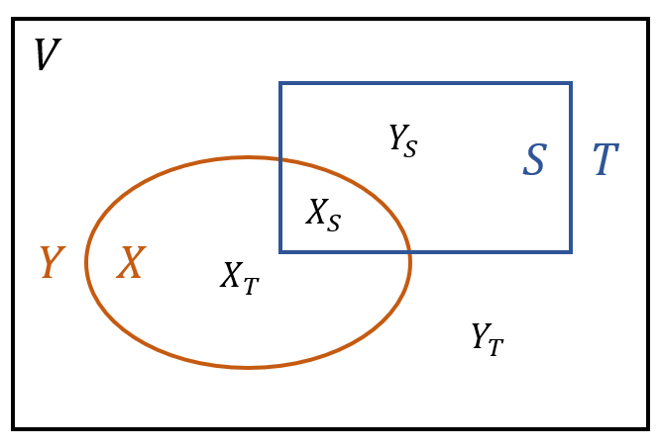}
	\caption{Illustration of the vertex subsets in \cref{eq:energy_criteria}. Here $X$ denotes the candidate for NG under consideration, $Y$ denotes its complement $Y:=V \backslash X$; $S$ is an arbitrary set that separates $X$, and $T:=V \backslash S$ is its complement.}
	\label{fig:XYST}
\end{figure}

Consider a set of spins $X$ as a candidate for a non-separable group. Suppose we have chosen a specific configuration $S \in \sep(X)$ for the first term in \cref{eq:nu_def}. Then for the second term, instead of taking the minimum over all configurations that do not separate $X$, we may consider four special cases (see \cref{fig:XYST}). (1) Excluding all spins in $X$ from $S$. We call such subset $Y_S$. (2) Including all spins except for those in $X$ and $S$ (complement of $X \cup S$). We call such configuration $Y_T$. (3) A complement of (1). (4) A complement of (2). Note that all these four configurations do not separate $X$, i. e. either all spins in $X$ are negative or all spins in $X$ are positive.
Then \cref{eq:nu_def} will look as follows
\begin{equation}\label{eq:energy_criteria}
	\nu_H(X) \geq  \frac{1}{2}\min_{S \in \sep(X)} \left[ \right. H(S) - \min \left( H(Y_S), H(Y_T),H(\bar{Y}_S), H(\bar{Y}_T)\left. \right)\right].
\end{equation}
$H(Y_S)$,$H(Y_T)$ correspond to the cases where all spins in $X$ take value $+1$, and $H(\bar{Y}_S)$, $H(\bar{Y}_T)$ correspond to the cases where all spins in $X$ take value $-1$. This step reduces the complexity of taking the minimum over all configurations that do not separate $X$ to four simple extensions of $S$.

After this step, the further derivation becomes quite complicated, so we will describe only the basic ideas behind it while leaving the technical details to Appendix~\cref{app:criteria}.

First, we rewrite~\cref{eq:energy_criteria} in an equivalent form:
\begin{equation}\label{eq:energy_criteria_nui_exact}
	\nu_H(X)  \geq \min_{S \in \sep(X)}\max_{1\leq i \leq 6} \nu_i(X;X_S),
\end{equation}
where
\begin{equation} \label{eq:energy_criteria_nui_def}
\begin{aligned}
    \nu_1(X;X_S) & = \frac{1}{2}\left(H(S)-H(Y_S) \right),  \\
    \nu_2(X;X_S) & = \frac{1}{2}\left(H(S)- H(\bar{Y}_T) \right),  \\
    \nu_3(X;X_S) & = \frac{1}{4}\left(2H(S)-H(Y_S)-H(\bar{Y}_T)\right),
\end{aligned}
\qquad
\begin{aligned}
    \nu_4(X;X_S) & = \frac{1}{2}\left(H(S)-H(\bar{Y}_S) \right),  \\
    \nu_5(X;X_S) & = \frac{1}{2}\left(H(S)- H(Y_T) \right),  \\
    \nu_6(X;X_S) & = \frac{1}{4}\left(2H(S)-H(\bar{Y}_S)-H(Y_T)\right).
\end{aligned}
\end{equation}
Here, equations for $\nu_3$ and $\nu_6$ are obtained as averages of $(\nu_1, \ \nu_2)$ and $(\nu_4, \ \nu_5)$, respectively. They are redundant in this exact form, but provide different lower bound estimations in further derivation. Each quantity $\nu_1, \dots, \nu_6$ can be further decomposed to consider different orders of interactions separately, according to \cref{eq:Ham}, 
\begin{equation}
    \nu_i(X;X_S) = \sum_{m=1}^{M} \nu_i^{(m)}(X;X_S),
\end{equation}
where $\nu_i^{(m)}(X;X_S)$, $i=1\dots6$, are defined in a similar way to~\cref{eq:energy_criteria_nui_def}, but considering only interactions of order $m$ instead of full Hamiltonian:
\begin{equation}\label{eq:M_order_Ham}
	H^{(m)}(\mathbf{\sigma}) = \sum_{\{i_1,...,i_m\} \in E} J_{i_1,...,i_m}\sigma_{i_1}\cdots \sigma_{i_m}.
\end{equation}
Note that for even $m$ the value of $H^{(m)}(\mathbf{\sigma})$ does not change if all spins in the configuration change their signs simultaneously, so only values of $\nu_i^{(m)}$ at $i=1\dots3$ have to be computed in practice. For odd value of $m$ all six quantities $\nu_i^{(m)}$ at $i=1\dots6$ are, in general, independent.

The key insight to the efficiency of the method, as shown in Appendix~\cref{app:criteria}, is that we can efficiently compute lower bounds for all $\nu_i^{(m)}(X;X_S)$ by considering only different separations of $X$, avoiding minimization over all  $S \in \sep(X)$, required in~\cref{eq:energy_criteria}. It gives us estimation, as follows:
\begin{equation}\label{eq:energy_criteria_nui_bond}
	\nu_H(X)  \geq \min_{S \in \sep(X)}\max_{1\leq i \leq 6} \nu_i(X;X_S) \geq \hat{\nu}_H(X) = \min_{\emptyset\neq X_S \subsetneq X}\max_{1\leq i \leq 6} \hat{\nu}_i(X;X_S),
\end{equation}
where
\begin{equation}\label{eq:criteria}
	\hat{\nu}_i(X;X_S):= \sum_{m=1}^{M}  \hat{\nu}_i^{(m)}(X;X_S),
\end{equation}
and $\nu_i^{(m)}(X;X_S) \geq \hat{\nu}_i^{(m)}(X;X_S)$ for all $m = 1 \dots M$.

For even $m$, $\hat{\nu}_{i+3}^{(m)}(X;X_S)=\hat{\nu}_i^{(m)}(X;X_S)$ for $i=1,2,3$ due to $Z_2$ symmetry. Detailed derivation and the explicit expression of lower bounds in \cref{eq:criteria} can be found in Appendix~\cref{app:criteria}.

This criterion is the foundation of our GeneralHare algorithm. If $\hat{\nu}_{H}(X)>0$, we identify $X$ as a non-separable group. While $\nu_{H}(X)$ involves the weights of all edges in $E$, when calculating $\hat{\nu}_H(X)$, we only need to consider the edges that link nodes inside $X$ and those in the neighborhood of $X$. This means that we can determine a generalized non-separable group by observing only the local information around it, resulting in a substantial reduction in computational cost.

In the case of $\hat{\nu}_{H}(X)=0$, set $X$ may be either non-separable or weakly non-separable, due to heuristic nature of the lower bound estimation. In this case we conservatively call the group weakly non-separable and process it accordingly. The only practical implication is that weakly non-separable groups should not be merged together if their neighborhoods intersect with each other. So, in this case we greedily construct an independent set of all groups that are detected as potentially being weakly non-separable, considering them together with their immediate neighborhoods, and apply contraction only to this set. Thus, if a non-separable group was misclassified as weakly non-separable, it may result in degraded reduction ratio, while the exact correspondence between ground states of the initial and the reduced models is still preserved.

\subsubsection{Generalized non-separable groups} 
For the purpose of discussion, we need a generalization on the concept of NGs. Consider a simple case where $X$ consists of only $2$ nodes. Apart from the case where $X$ is a (weakly) NG, it is possible that the two spins in $X$ take opposite signs in all ground states, in which case they are called an `antipolar pair'~\cite{thai2022fasthare}. As the size of $X$ increases, more possible partial configurations within $X$ may emerge, which calls for a generalized version of NG. The basics properties of NGs are still valid for gNGs.

A set $X$ is called a generalized non-separable group (gNG) if there exists a partial configuration $\sigma_X \in \{-1,+1\}^{|X|}$ ($|X|$ is the size of $X$), such that for all ground states, the spins inside $X$ take the configuration of either $\sigma_X$ or $-\sigma_X$, and the weakly generalized non-separable group (weakly gNG) is defined similarly. As a result, all spins in a (weakly) gNG can be merged into a single spin, creating a reduced Hamiltonian of a smaller size with a reconstruction map preserving ground-state correspondence. Since the merged spin has one degree of freedom, only the `relative’ configuration inside $X$ matters. To remove ambiguity, we choose $\sigma_X$ such that the spin with the smallest index in $X$ is $+1$.

By flipping the nodes that take $-1$ in $\sigma_X$, we can convert a gNG into an NG. The flipping of a node is done by negating the weights of all hyperedges incident at it (its linear bias is considered as a hyperedge of order 1). Concretely, we have the following proposition. 
\begin{proposition}\label{prop:gNG}
	Suppose $Z \subseteq X$ consists of all nodes that take $-1$ in $\sigma_X$. Define
	\begin{equation}
		\tilde{H}(\mathbf{\sigma}) = \sum_{I\in E} (-1)^{n_I(Z)} J_I \cdot \left(\prod_{i\in I}\sigma_i\right).
	\end{equation}
	Then $X \subseteq V$ is a gNG for Hamiltonian $H$ with relative configuration $\mathbf{\sigma}_X$ if and only if $X$ is a non-separable group for Hamiltonian $\tilde{H}$.
\end{proposition}

Thus iterating over all partial configurations $\mathbf{\sigma}_X$ of $X$, changing $\tilde{H}$ accordingly, and calculating $\nu_{\tilde{H}}(X)$ would allow us to determine whether $X$ forms a (weakly) gNG. Note that for a problem that has single non-degenerate ground state, or two ground states with $Z2$ symmetry, the whole set of nodes by definition forms a gNG. In this case, detecting gNGs exactly is equivalent to solving the problem. In practice, we compute lower bounds $\hat{\nu}_{\tilde{H}}(X)$, thus detecting (in general) only a fraction of gNGs and candidates for weakly gNGs, but in a more efficient way which is critical for practical applications.

\subsection{The GeneralHare scheme}\label{sec:algorithm}

As illustrated in \cref{fig:schematic}, GeneralHare consists mainly of three stages: node fixation, identification of (weakly) gNGs, and enlarging and compression of gNGs. The procedure will be repeated for multiple rounds until no more gNGs can be found. GeneralHare serves as a preprocessing step for the original Hamiltonian. It outputs a smaller reduced Hamiltonian together with the information needed to reconstruct original spin configurations from reduced solutions. In this way, the GeneralHare will lead to a reduction on quantum (for generic quantum devices) or classical (for classical or quantum-inspired solvers) computational resources.

\subsubsection{Node fixation} 
In the second-order case, the linear terms are resolved with the ancilla spin through SK graph, which is no longer valid now due to the absence of $Z_2$ symmetry of general-order Hamiltonians. To deal with linear terms in every iteration, we introduce a step called `node fixation' before identifying gNGs. For a node $u$, if the absolute value of its linear term is larger than the sum of the absolute values of weights of all hyperedges (of order $m\geq 2$) attached to this node, i.e., 
\begin{equation}
	|J_u|> \sum_{\substack{I\in E \ \text{s.t.}\\ |I|\geq2, u\in I}} |J_I|,
\end{equation}
then we can fix the spin of node $u$. If $J_u>0$, then $\sigma_u=-1$, otherwise $\sigma_u = +1$. 

After fixing \(\sigma_u=s_u\), each incident term
\(J_I\prod_{i\in I}\sigma_i\) with \(u\in I\) is replaced by
\[
    J_I s_u \prod_{i\in I\setminus\{u\}}\sigma_i .
\]
Thus fixation can also produce lower-order terms, and the corresponding linear term $J_us_u$ becomes an additive constant. As in gNG compression~(\cref{sec:gng_compression}), this constant may be tracked explicitly or omitted if energies are later evaluated after reconstructing the original configuration.

\subsubsection{Identification of gNGs} 
In this stage, we identify gNGs based on Proposition~\ref{prop:gNG} and the criteria we derived. For a hyperedge $X$, to determine whether it forms a gNG, a naive approach would involve a computational cost of approximately $4^{|X|}$; this is the most compute-intensive part of the GeneralHare scheme. To reduce the cost, we introduce an integer $\xi$ ($\xi \geq 2$) to restrict the size of gNG candidates. Specifically, we choose gNG candidates from the hyperedges of order less than or equal to $\xi$, and with nonzero weight. This choice may leave some gNGs undetected, especially in problems with sparse or weak second-order interactions. In certain cases selecting higher values of $\xi$ may be beneficial, but our experiments on synthetic random hypergraphs show that $\xi=2$ provides a good trade-off between reduction ratio and speed. See Appendix~\cref{app:xi_sweep} for details.

Moreover, since the calculation of $\hat{\nu}_{\tilde{H}}$ involves taking the minimum over all bisections, we can do a pruning to further speed up the algorithm. For a relative configuration $\mathbf{\sigma}_X$, as soon as we encounter a bisection $X_S, X\backslash X_S$ such that $\hat{\nu}_{\tilde{H}}(X;X_S)<0$, we can immediately discard this configuration. The pseudo code is shown in Algorithm~\ref{alg:find_NGs}.

\begin{algorithm}
\caption{Find gNGs}\label{alg:find_NGs}
\begin{algorithmic}[1]
\Require A hypergraph $G=(V,E,J)$, an integer $\xi$, node list $\Omega$
\Ensure A list of gNGs $\Upsilon_s$, a list of weakly gNGs $\Upsilon_w$

\For{$X \in E$ such that $|X|\leq \xi$, $J_X \neq 0$, $X \cap \Omega \neq \emptyset$}
    \For{$\mathbf{\sigma}_X \in \{-1,+1\}^{|X|}$}
        \For{$X_S \subseteq X$ ($X_S \notin \{\emptyset, X\}$)}
            \State Calculate $\hat{\nu}_{\tilde{H}}(X;X_S)$
            \If{$\hat{\nu}_{\tilde{H}}(X;X_S)<0$}
                \State Discard $\mathbf{\sigma}_X$
            \EndIf
        \EndFor
        \State $\hat{\nu}_{\tilde{H}}(X)=\min_{X_S}\hat{\nu}_{\tilde{H}}(X;X_S)$
        \If{$\hat{\nu}_{\tilde{H}}(X)>0$}
            \State Append $X$ with config $\mathbf{\sigma}_X$ to $\Upsilon_s$
        \EndIf
        \If{$\hat{\nu}_{\tilde{H}}(X)=0$}
            \State Append $X$ with config $\mathbf{\sigma}_X$ to $\Upsilon_w$
        \EndIf
    \EndFor
\EndFor
\end{algorithmic}
\end{algorithm}

\subsubsection{Enlarging of gNGs} 
According to the `closure under union' property, now we can integrate the gNGs identified in the previous stage. We denote the partial configuration of $Z$ in $\mathbf{\sigma}_X$ as $\mathbf{\sigma}_X(Z)$ if $Z \subseteq X$. Once we find two gNGs $X$, $Y$, with relative configurations $\mathbf{\sigma}_X$ and $\mathbf{\sigma}_Y$, respectively, then $Z=X \cap Y$ is also a gNG, which means that it keeps its relative configuration in all ground states. Then we have either $\mathbf{\sigma}_X(Z)=\mathbf{\sigma}_Y(Z)$, or $\mathbf{\sigma}_X(Z)=-\mathbf{\sigma}_Y(Z)$.
	
If $\mathbf{\sigma}_X(Z)=\mathbf{\sigma}_Y(Z)$, $X \cup Y$ is a gNG with relative configuration $\mathbf{\sigma}_X \oplus \mathbf{\sigma}_Y$, where $\oplus$ denotes concatenation, that is, for any $i \in X\cup Y$, if $i \in X$, then $\mathbf{\sigma}_{X\cup Y}(\{i\}) = \mathbf{\sigma}_X(\{i\})$, otherwise $\mathbf{\sigma}_{X\cup Y}(\{i\}) = \mathbf{\sigma}_Y(\{i\})$. On the other hand, if $\mathbf{\sigma}_X(Z)=-\mathbf{\sigma}_Y(Z)$, then $X \cup Y$ is a gNG with relative configuration $\mathbf{\sigma}_X \oplus \left(-\mathbf{\sigma}_Y \right)$. 

\begin{algorithm}[t]
\caption{GeneralHare}\label{alg:GH}
\begin{algorithmic}[1]
\Require A hypergraph $G=(V,E,J)$, an integer $\xi$
\Ensure A reduced hypergraph $G$
\State Set $\Omega=V$
\While{True}
    \State Node fixation
    \State $\Upsilon_s, \Upsilon_w =$ Find gNGs($G,\xi,\Omega$)
    \If{$\Upsilon_s$, $\Upsilon_w$ are both $\emptyset$}
        \State Return $G$
    \Else
        \If{$\Upsilon_s \neq \emptyset$}
            \State Merge the gNGs in $\Upsilon_s$ to obtain a new list $\Upsilon$
        \Else
            \State Take an independent set of $\Upsilon_w$ as $\Upsilon$
        \EndIf
        \State Update $G$ according to $\Upsilon$. Obtain $\Omega$
    \EndIf
\EndWhile
\end{algorithmic}
\end{algorithm}
    
\subsubsection{Compression of gNGs} 
\label{sec:gng_compression}
In the next step, we compress a gNG (or one detected as a weakly gNG) $X$ into a single node. Suppose the relative configuration of $X$ is $\mathbf{\sigma}_X$, and $Z$ is a subset of $X$ that contains all nodes that take $-1$ in $\mathbf{\sigma}_X$. First, for every node in $Z$, we need to flip it by negating the weights of all hyperedges incident at it. Secondly, we merge all nodes in $X$. Suppose the hypergraph before and after the merging of $X$ are $G$ and $G_c$, respectively, and $X$ is merged into node $x$. For any hyperedge $I$ in $G$, its corresponding $I_c$ in $G_c$ reads
\begin{equation}
    I_c = 
    \begin{cases}
        I\backslash X \quad \text{if} \ n_I(X)\!:\!\text{EV}, \\
        \{x\} \cup I\backslash X \quad \text{if} \ n_I(X)\!:\!\text{OD}, \\
    \end{cases}
\end{equation}
and the weights of parallel edges also need to be aggregated.

It is important to distinguish the reduced objective from the energy value reported for a reconstructed solution. If \(I_c=\emptyset\), the corresponding term becomes an additive constant after compression. More generally, compression may also turn higher-order terms into lower-order terms. Additive constants do not affect the minimizers of the reduced Hamiltonian, so they may either be tracked explicitly or omitted during the reduction. If the accumulated constant is denoted by \(C\), then the reconstruction map \(R\) from reduced configurations to original configurations satisfies
\begin{equation}
    H(R(\boldsymbol{\tau})) = H_c(\boldsymbol{\tau}) + C ,
\end{equation}
for every reduced configuration \(\boldsymbol{\tau}\), whenever \(C\) is explicitly tracked. If \(C\) is not tracked, the same reduced Hamiltonian can still be used to identify minimizers, but absolute reduced energy values should not be compared directly with the original Hamiltonian. In our solver-level evaluations, energies are therefore reported after reconstructing the original spin configuration and evaluating it on the original Hamiltonian.

\subsubsection{Speed up the procedure} 
We make two optimizations to further accelerate the search for and merging of (weakly) gNGs. First, we can keep a list of nodes, denoted as $\Omega$, which contains all newly updated nodes and their neighborhoods in the last iteration (initially, $\Omega=V$). Evidently, for hyperedge $X$ that does not intersect with $\Omega$, $\hat{\nu}_{\hat{H}}(X)$ is unchanged from the last iteration, meaning that $\hat{\nu}_{\hat{H}}(X)<0$ for all $\mathbf{\sigma}_X$. So we can exclude gNG candidates that are not in $\Omega$ from our consideration to reduce computational cost. 

Second, the procedure of merging weakly gNGs can also be optimized. In FastHare, only one weakly NG is merged in every iteration, which turns out to be inefficient due to the large number of node sets detected as weakly gNGs for hypergraphs. To speed up the reduction, we develop the technique of simultaneous merging of weakly gNGs. 

For those groups detected as weakly gNGs, we merge several groups in the same iteration only when their closed hypergraph neighborhoods are disjoint. Under this condition, the corresponding contractions do not interfere with each other and the simultaneous step is equivalent to applying the same weak-gNG contractions sequentially. We therefore greedily construct an approximate maximal independent set of weakly gNGs and contract only this set in one iteration.

So far, we have explained the whole procedure of our scheme in detail, and the pseudo code of GeneralHare is presented as Algorithm~\ref{alg:GH}, where the two techniques described above are shown in lines 13 and 11, respectively.

\section{Discussion}

We generalized the concept of non-separable groups and the theory of non-separability, based on which we proposed the GeneralHare scheme for reducing the size of an Ising-like Hamiltonian. GeneralHare reduces the number of logical spin variables in higher-order Ising-like Hamiltonians and supporting solver-level experiments suggest that reduction can improve heuristic solver performance on the tested instances. Our work extends the horizon of Hamiltonian reduction to higher-order Ising-like problems, which arise naturally in many pseudo-Boolean formulations and higher-order network models. However, as a generally applicable scheme, the effect of GeneralHare may diminish on certain classes of problems, such as $3$-SAT problems whose associated hypergraphs have homogeneous degree distributions. It would be of practical interest if a Hamiltonian reduction scheme designed specifically for this problem could be developed. One may consider combining Hamiltonian reduction with classical SAT solvers such as kissat~\cite{Biere2020}, and exploring the deeper connection between the underlying ideas.
We believe that our work can inspire further advances in both research and applications of Ising-like models.

\subsection*{Data availability}
\addcontentsline{toc}{section}{Data and code availability}

The public higher-order network datasets used in this work are available from
Ref.~\cite{benson2018}. The synthetic instances were generated according to the
procedures described in Appendix~\cref{app:data}. The specific random instances,
random seeds, processed data files, and scripts used in the original numerical
experiments were not retained in an accessible repository and are therefore not
available.

\subsection*{Code availability}

Two implementations were used in this work. The first is the original full implementation of GeneralHare in C++, which includes the non-separability criteria based on the efficient lower bounds in~\cref{eq:criteria}, gNG enlargement, tracking of merged nodes, and optimized processing of variable sets identified as weakly gNGs. This implementation was used for the main performance experiments in~\cref{sec:performance} and~\cref{sec:fh_comparison}, as well as for several evaluations reported in the Appendix. The full C++ implementation is not available for public release.

To support reproducibility and independent inspection of the core reduction mechanism, we provide a separate minimal Python reference implementation, denoted GH\_minimal, available at~\url{https://github.com/PMosharev/GH_minimal}. GH\_minimal implements a conservative version of the certified reduction procedure. Instead of the full efficient lower-bound criteria in~\cref{eq:criteria}, it uses exhaustive configuration search over the immediate neighborhood of each candidate group. It also omits gNG enlargement, simultaneous merging of multiple gNGs, detection and optimized handling of weakly gNG candidates, and other optimizations. This implementation is intended as a transparent conceptual demonstrator rather than a replacement for the full C++ code. GH\_minimal was used for the PyQUBO/order-reduction experiment in~\cref{sec:pyqubo_or_combination} and for additional reproducibility demonstrations in the Appendix.

The original C++ implementation follows a FastHare-style convention in which additive constants generated during compression are not used for reporting absolute reduced energies. In solver-level experiments, reduced solutions are reconstructed and evaluated on the original Hamiltonian. The released GH\_minimal implementation additionally tracks constant offsets explicitly to make the algebraic compression and validation tests transparent.

\subsection*{Acknowledgements}
This work is supported by Project 12047503, 12325501, and 12247104 of the National Natural Science Foundation of China and project ZDRW-XX-2022-3-02 of the Chinese Academy of Sciences. 

\subsection*{Author contribution}
C.M., P.M. and Y.W. conceived the idea,  C.M. and P.M. wrote the codes and conducted the experiments, C.M. and Y.W. analyzed the results. M.-H.Y coordinated and supervised the work. All authors reviewed the manuscript. 

\bibliographystyle{unsrtnat}
\bibliography{ref}

\begingroup
\scriptsize 
\tableofcontents
\endgroup

\clearpage

\section*{APPENDIX}
\addcontentsline{toc}{section}{APPENDIX} 
\appendix
\appendix

In the Appendix, we give a detailed derivation of the criteria for non-separability we use in the main text, the
algorithm we use for synthesizing datasets, a brief introduction to high-order simulated bifurcation, as well as some
results from additional numerical experiments.

\section{Detailed derivation of criteria for non-separability}
\label{app:criteria}

We recall the notation used in the main text. Let \(V=[n]\), and let
\(\sigma_i\in\{-1,+1\}\) be the spin associated with node \(i\). An
order-\(M\) Ising-like Hamiltonian is written as
\begin{equation}
\label{eq:supp_ham}
H(\boldsymbol{\sigma}) = \sum_{\emptyset\neq I\subseteq V,\, |I|\le M} J_I\prod_{i\in I}\sigma_i ,
\end{equation}
where each hyperedge \(I\) appears once. For a subset \(S\subseteq V\), we
use the equivalent set notation in which \(\sigma_i=-1\) for \(i\in S\) and
\(\sigma_i=+1\) for \(i\notin S\). For a hyperedge \(I\), define
\begin{equation}
\label{eq:supp_nI}
n_I(S):=|I\cap S|.
\end{equation}
Then
\begin{equation}
\label{eq:supp_HS}
H(S) = \sum_{n_I(S):\mathrm{EV}} J_I - \sum_{n_I(S):\mathrm{OD}} J_I = \sum_{I\in E}J_I - 2\sum_{n_I(S):\mathrm{OD}}J_I .
\end{equation}

For \(X\subseteq V\), a configuration \(S\) separates \(X\), denoted
\(S\ominus X\), if
\begin{equation}
\label{eq:supp_sep_condition}
X\cap S\notin\{\emptyset,X\}.
\end{equation}
We write
\begin{equation}
\label{eq:supp_sep}
\sep(X):=\{S\subseteq V:\, S\ominus X\}.
\end{equation}
The non-separability index of \(X\) is
\begin{equation}
\label{eq:supp_nu}
\nu_H(X)
=
\frac{1}{2}
\left[
\min_{S\in\sep(X)}H(S)
-
\min_{S'\notin\sep(X)}H(S')
\right].
\end{equation}

For \(S\in\sep(X)\), let $Y:=V\setminus X,\quad T:=V\setminus S$,
and define $X_S:=X\cap S,\quad X_T:=X\cap T,\quad Y_S:=Y\cap S,\quad Y_T:=Y\cap T$. 
The derivation below uses the following lower bound:
\begin{equation}
\label{eq:supp_energy_criteria}
\nu_H(X)
\geq
\frac{1}{2}
\min_{S\in\sep(X)}
\left[
H(S)
-
\min\left(
H(Y_S),H(Y_T),H(\bar{Y}_S),H(\bar{Y}_T)
\right)
\right].
\end{equation}

Finally, the \(m\)-th order part of the Hamiltonian is
\begin{equation}
\label{eq:supp_Hm}
H^{(m)}(\boldsymbol{\sigma})
=
\sum_{\substack{I\in E\\ |I|=m}}
J_I\prod_{i\in I}\sigma_i .
\end{equation}
In set notation,
\begin{equation}
\label{eq:supp_Hm_set}
H^{(m)}(S)
=
\sum_{\substack{I\in E\\ |I|=m}}
(-1)^{n_I(S)}J_I .
\end{equation}
Therefore, for \(\bar S:=V\setminus S\),
\begin{equation}
\label{eq:supp_parity}
H^{(m)}(\bar S)
=
\begin{cases}
H^{(m)}(S), & m \text{ even},\\
-H^{(m)}(S), & m \text{ odd}.
\end{cases}
\end{equation}

Before deriving bounds for $\nu_H(X)$, we introduce some necessary notations. Let $H$ be an Ising-like Hamiltonian with highest order $M$, and $G:=(V,E,J)$ be its corresponding hypergraph. The $m$-th-order cut $c(X_1,X_2,...,X_m)$ is defined as a sum of weights of all $m$th-order hyperedges across parities $X_1,X_2,...,X_m$, and $c_{|\cdot|}(X_1,X_2,...,X_m)$ is defined similarly, but with all terms taking the absolute value

\begin{align}
	c(X_1,X_2,...,X_m) & := \sum_{i_1\in X_1,i_2\in X_2,...,i_m \in X_m} J_{i_1,i_2,...,i_m}, \nonumber\\
	c_{|\cdot|}(X_1,X_2,...,X_m) & := \sum_{i_1\in X_1,i_2\in X_2,...,i_m \in X_m} |J_{i_1,i_2,...,i_m}|.
\end{align}
When $G$ is a (second-order) SK graph, $c(X_1,X_2)$ and $c_{|\cdot|}(X_1,X_2)$ recovers the cuts defined in FastHare \cite{thai2022fasthare}. 

To simplify the notation, we also define the $m$th-order `cut sum' $C^{(m)}[X_1,X_2,...,X_k\,|\, (\text{conditions})]$, which equals to the sum of all $c(X_1,X_2,...,X_m)$ such that $X_1,X_2,...,X_m$ is a collection of $X_1,X_2,...,X_k$ ($k \leq m \leq M$), and the conditions are satisfied, where $X_1,X_2,...,X_k$ must all be different, but elements in $X_1,X_2,...,X_m$ may repeat.

For instance, consider a Hamiltonian with $M=5$. Let $P = \{0,1\}, Q=\{2,3,4\}, R = \{5\}$ be three subsets of $V$, then
\begin{align}
	& C^{(4)}[P,Q,R \,|\, n(P)\!:\! \text{EV}, n(Q)\!:\!\text{OD}, n(R)\neq 0] \nonumber\\
	= & c(P,P,Q,R)+c(Q,Q,Q,R) \nonumber\\
	= & J_{0125}+J_{0135}+J_{0145}+J_{2345}.
\end{align}
Here we omit the subscript $I$ in $n_I(\cdot)$ for simplicity, keeping in mind that for each term in the sum $n(S)$ corresponds to the hyperedge. 
$C^{(m)}_{|\cdot|}[X_1,X_2,...,X_k\,|\, (\text{conditions})]$ is defined similarly.

Then we can derive the explicit expression of the RHS of \cref{eq:supp_energy_criteria} for non-separability index
for even and odd orders, respectively, and sum them up from $1$ to $M$ to derive a lower bound for $\nu_H(X)$.

When $m$ is an even number, we have
\begin{align}\label{eq:energy_criteria_m}
	\nu_1^{(m)}(X;X_S) & = \frac{1}{2}\left(H^{(m)}(S)-H^{(m)}(Y_S)\right) = -\sum_{n(S):\text{OD}}J_I + \sum_{n(Y_S):\text{OD}} J_I\nonumber\\
	& \geq -C^{(m)}\left[X_S,X_T \,|\, n(X_S)\!:\!\text{OD},n(X_T)\!:\!\text{OD}\right]\nonumber\\
	& \quad - C^{(m)}_{|\cdot|}\left[X_S,X_T,Y\,|\, n(X_S)\!:\!\text{OD},n(Y)\neq 0\right]\nonumber\\
	&=:\hat{\nu}_1^{(m)}(X;X_S),
\end{align}
and similarly, 
\begin{align}
	\nu_2^{(m)}(X;X_S) & = \frac{1}{2}\left(H^{(m)}(S)-H^{(m)}(\bar{Y}_T)\right) \nonumber\\
	&\geq -C^{(m)}\left[X_S,X_T \,|\, n(X_S)\!:\!\text{OD},n(X_T)\!:\! \text{OD}\right] \nonumber\\
	& \quad- C^{(m)}_{|\cdot|}\left[X_S,X_T,Y\,|\, n(X_T)\!:\! \text{OD},n(Y)\neq 0\right]\nonumber\\
	&=:\hat{\nu}_2^{(m)}(X;X_S).
\end{align}
\begin{align}\label{eq:nuhat3m}
	\nu_3^{(m)}(X;X_S) & = \frac{1}{4}\left(2H^{(m)}(S)-H^{(m)}(Y_S)-H^{(m)}(\bar{Y}_T)\right) \nonumber\\
	& \geq  -C^{(m)}\left[X_S,X_T \,|\, n(X_S)\!:\!\text{OD},n(X_T)\!:\!\text{OD}\right]\nonumber\\
	& \quad - C^{(m)}_{|\cdot|}\left[X_S,X_T,Y \,|\, n(X_S)\!:\!\text{OD},n(X_T)\!:\!\text{OD},n(Y)\!:\!\text{EV},n(Y)\neq 0\right]\nonumber\\
	& \quad - \sum_{\substack{i_1,...,i_{m-1} \text{s.t.} \\ n(X_S;I\backslash\{i_m\}):\text{EV} \\ n(X_T;I\backslash\{i_m\}):\text{EV} \\ n(Y;I\backslash\{i_m\}):\text{OD}}} \lvert\sum_{i_m\in X_S} J_I- \sum_{i_m\in X_T}J_I\rvert \nonumber\\
	& =: \hat{\nu}_3^{(m)}(X;X_S).
\end{align}

We explain the third term in \cref{eq:nuhat3m} a little more.
Since any hyperedge $I=\{i_1,...,i_m\} \in E$ corresponds to a unique interaction term $J_I$ for a given Hamiltonian, for simplicity of expression, we permute the indices such that all $J_I$ appear inside the absolute value as the same index set except the last one $i_m$. 

When \(m\) is odd, the same lower-bound strategy applied to \cref{eq:supp_energy_criteria} gives the following three bounds:
\begin{align}
	\hat{\nu}_1^{(m)}(X;X_S) &:= -C^{(m)}\left[X_S,X_T \,|\, n(X_S)\!:\!\text{OD},n(X_T)\!:\!\text{EV}\right] - C^{(m)}_{|\cdot|}\left[X_S,X_T,Y \,|\, n(X_S)\!:\!\text{OD},n(Y)\neq 0\right], \\
	\hat{\nu}_2^{(m)}(X;X_S) &:=  C^{(m)}\left[X_S,X_T \,|\, n(X_S)\!:\!\text{EV},n(X_T)\!:\!\text{OD}\right] - C^{(m)}_{|\cdot|}\left[X_S,X_T,Y \,|\, n(X_T)\!:\!\text{OD},n(Y)\neq 0\right], \\
	\hat{\nu}_3^{(m)}(X;X_S) &:=\frac{1}{2} C^{(m)}\left[X_S,X_T \,|\, n(X_S)\!:\!\text{EV},n(X_T)\!:\!\text{OD}\right]-\frac{1}{2} C^{(m)}\left[X_S,X_T \,|\, n(X_S)\!:\!\text{OD},n(X_T)\!:\!\text{EV}\right]\nonumber\\ 
	& \quad - C^{(m)}_{|\cdot|}\left[X_S,X_T,Y \,|\, n(X_S)\!:\!\text{OD},n(X_T)\!:\!\text{OD},n(Y)\!:\!\text{OD}\right]\nonumber\\
	& \quad - \sum_{\substack{i_1,...,i_{m-1} \text{s.t.} \\ n(X_S;I\backslash\{i_m\}):\text{EV} \\ n(X_T;I\backslash\{i_m\}):\text{EV} \\ n(Y;I\backslash\{i_m\}):\text{EV} \\ n(Y;I\backslash\{i_m\})\neq 0 }} \lvert\sum_{i_m\in X_S} J_I- \sum_{i_m\in X_T}J_I\rvert .
\end{align}

For odd $m$, however, besides the three criteria we just derived, we also need to take the condition 
\begin{equation}
	\nu_{H^{(m)}}(X) \geq \frac{1}{2}\min_{S \in sep(X)}\left[ H^{(m)}(S)- \min \left(  H^{(m)}(\bar{Y}_S),H^{(m)}(Y_T)\right)\right]
\end{equation}
into consideration, which leads to three extra lower bounds on $\nu_{H^{(m)}}(X)$:
\begin{align}
	\hat{\nu}_4^{(m)}(X;X_S) &:=  C^{(m)}\left[X_S,X_T \,|\, n(X_S)\!:\!\text{EV},n(X_T)\!:\!\text{OD}\right] - C^{(m)}_{|\cdot|}\left[X_S,X_T,Y \,|\, n(X_S)\!:\!\text{EV},n(Y)\neq 0\right], \\
	\hat{\nu}_5^{(m)}(X;X_S) &:=  -C^{(m)}\left[X_S,X_T \,|\, n(X_S)\!:\!\text{OD},n(X_T)\!:\!\text{EV}\right] - C^{(m)}_{|\cdot|}\left[X_S,X_T,Y \,|\, n(X_T)\!:\!\text{EV},n(Y)\neq 0\right], \\
	\hat{\nu}_6^{(m)}(X;X_S) &:=\frac{1}{2} C^{(m)}\left[X_S,X_T \,|\, n(X_S)\!:\!\text{EV},n(X_T)\!:\!\text{OD}\right] -\frac{1}{2} C^{(m)}\left[X_S,X_T \,|\, n(X_S)\!:\!\text{OD},n(X_T)\!:\!\text{EV}\right]\nonumber\\ 
	& \quad - C^{(m)}_{|\cdot|}\left[X_S,X_T,Y \,|\, n(X_S)\!:\!\text{EV},n(X_T)\!:\!\text{EV},n(Y)\!:\!\text{OD}\right]\nonumber\\
	& \quad - \sum_{\substack{i_1,...,i_{m-1} \text{s.t.} \\ n(X_S;I\backslash\{i_m\}):\text{OD} \\ n(X_T;I\backslash\{i_m\}):\text{OD} \\ n(Y;I\backslash\{i_m\}):\text{EV} \\ n(Y;I\backslash\{i_m\})\neq 0 }} \lvert\sum_{i_m\in X_S} J_I- \sum_{i_m\in X_T}J_I\rvert .
\end{align}
For even $m$, by definition, $\hat{\nu}_{i+3}^{(m)}(X;X_S)=\hat{\nu}_i^{(m)}(X;X_S)$ for $i=1,2,3$. Finally, we get the lower bounds:
\begin{equation}
\nu_H(X)  \geq \hat{\nu}_H(X) = \min_{\emptyset\neq X_S \subsetneq X}\max_{1\leq i \leq 6} \hat{\nu}_i(X;X_S) = \min_{\emptyset\neq X_S \subsetneq X}\max_{1\leq i \leq 6} \left( \sum_{m=1}^{M}  \hat{\nu}_i^{(m)}(X;X_S) \right).
\end{equation}

If $\hat{\nu}_{H}(X)>0$, we identify $X$ as a non-separable group. In case $\hat{\nu}_{H}(X)=0$, the criterion is inconclusive, because \(\hat{\nu}_{H}\) is only a lower bound on \(\nu_H\). In this case we conservatively treat $X$ as weakly non-separable and process it correspondingly.

Note that $X_S$, $X_T$ ($X_S, X_T \notin \{\emptyset, X\}$) actually determine a bisection of $X$. According to \cref{eq:supp_energy_criteria}, after taking the summation of all order terms, we should do minimization over all $S\in \sep(X)$, that is, minimizing over all bisection of $X$. Furthermore, we can replace $Y$ with the neighborhood of $X$ (a subset of $V\backslash X$ containing all nodes that links to one or more nodes in $X$ with at least one hyperedges), since otherwise the corresponding $J_I$ would be $0$ and makes no contribution. This means that we can determine a non-separable group by observing only the local information around it, which is favorable in terms of computational complexity.

\section{Details on dataset}
\label{app:data}
\subsection{Generation of synthesized data}
In this section, we describe the synthetic hypergraph generators used in the numerical experiments and in the released reference implementation. We consider three families of synthetic hypergraphs. The ER-like generator produces sparse hypergraphs with approximately homogeneous degree distribution. The SF-like generator uses preferential attachment to produce heavy-tailed degree distributions. The regular-local generator starts from a bounded-degree regular graph and samples higher-order interactions from local one-hop neighborhoods; it is used to test the bounded-locality regime in which local certification is expected to be most effective.

For the ER-like and SF-like generators, we introduce parameters \(\{d_k\}_{k=2}^{M}\), where \(d_k\) denotes the target average degree per node for order-\(k\) interactions. Thus the target number of order-\(k\) hyperedges is approximately \(d_k n/k\). For the regular-local generator, the pairwise degree is controlled separately by the regular backbone degree \(d_{\mathrm{reg}}\), while \(\{d_k\}_{k=3}^{M}\) specifies target average degrees for higher-order local interactions.

\begin{figure}[tb]
	\centering
	\includegraphics[width=0.9\textwidth]{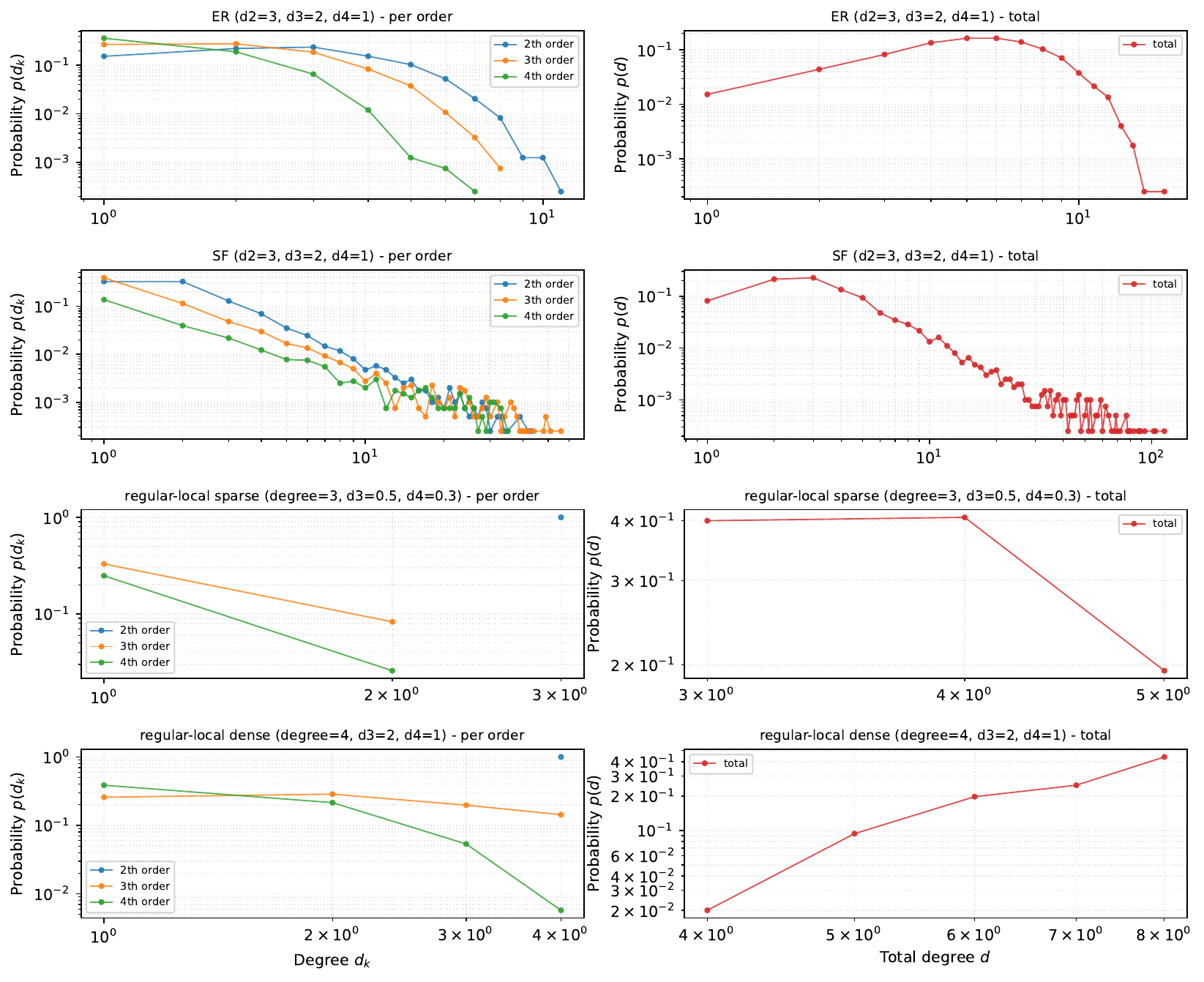}
	\caption{\textbf{Log-log plot of node degree versus its probability in a randomly generated ER/SF-like, as well as regular-local hypergraphs.} Each row corresponds to one generator setting, and the two columns show order-specific degree distributions \(p(d_k)\) and total-degree distributions \(p(d)\), respectively. The ER-like and SF-like hypergraphs use \(n=200\) nodes and target average degrees \(d_2=3\), \(d_3=2\), and \(d_4=1\). The regular-local examples use a random regular pairwise backbone and higher-order interactions sampled from closed one-hop neighborhoods: the sparse setting uses \(d_{\mathrm{reg}}=3\), \(d_3=0.5\), and \(d_4=0.3\), while the dense setting uses \(d_{\mathrm{reg}}=4\), \(d_3=2\), and \(d_4=1\). All instances are unweighted, and each curve is averaged over 20 independently generated samples. Zero-degree nodes are included in the normalization of the empirical probabilities but omitted from the log-log plots. 
    }
	\label{fig:ERSF}
\end{figure}

\textbf{ER-like hypergraph}: In this kind of hypergraph, each hyperedge has a fixed probability of being present or absent, independent of other hyperedges. \cref{alg:ER} describes the procedure of generating a random unweighted ER-like hypergraph. 

Denote the probability of a hyperedge of order $k$ to be present as $p_k$. In the sparse regime used in our experiments, duplicate rejection is rare, and the sampling strategy in \cref{alg:ER} is equivalent to setting 
\begin{equation}
	p_k = \frac{d_k}{\binom{n-1}{k-1}}
\end{equation}
for $k=2,...,M$, and iterating over all possible hyperedges to decide whether it exists according to $p_k$. We use \cref{alg:ER} to avoid iterating over all combinations, so the expected runtime is proportional to the number of sampled hyperedges, i.e. \(O(n\sum_k d_k)\). For very dense parameter choices, rejection sampling is no longer efficient and the implementation raises an error if too many duplicate attempts occur.

\begin{algorithm}[t]
\caption{Generating a random ER-like hypergraph}
\label{alg:ER}
\begin{algorithmic}[1]

\Require Number of nodes \(n\), highest order \(M\), average degrees \(\{d_k\}_{k=2}^{M}\)
\Ensure An unweighted hypergraph \(G=(V,E)\)

\State Set \(V\leftarrow\{1,\ldots,n\}\), \(E\leftarrow\emptyset\)

\For{\(k=2,\ldots,M\)}
    \If{\(d_k=0\)}
        \State \textbf{continue}
    \EndIf

    \State Set \(q_k \leftarrow d_k n/k\)
    \State Draw \(N_k \leftarrow \lfloor q_k\rfloor + \operatorname{Bernoulli}(q_k-\lfloor q_k\rfloor)\)
    \State Set \(\mathrm{cnt}=0\)

    \While{\(\mathrm{cnt}<N_k\)}
        \State Sample a \(k\)-element subset \(e\subseteq V\) uniformly at random
        \If{\(e\notin E\)}
            \State Add \(e\) to \(E\)
            \State \(\mathrm{cnt}\leftarrow \mathrm{cnt}+1\)
        \EndIf
    \EndWhile

\EndFor

\State \Return \(G=(V,E)\)

\end{algorithmic}
\end{algorithm}

Erd{\H{o}}s-R{\'e}nyi network has a deep relation with the percolation theory in physics and the evolution of a random network~\cite{erdHos1960evolution}. We expect ER-like hypergraphs to possess similar properties. The first row of~\cref{fig:ERSF} illustrates the degree distribution of a typical ER-like hypergraph.

\textbf{SF-like hypergraph}: This generator is a higher-order preferential-attachment model designed to produce heavy-tailed, scale-free-like degree distributions.
To generate random SF-like hypergraphs, we propose \cref{alg:sf-hypergraph}, which is a modified version of preferential attachment algorithm \cite{barabasi2002evolution}, i.e., the more one node is connected to others, the more likely it is chosen to build up new interactions.

Denote by $q_k$ the number of $k$-th order hyperedges a new node builds up with existing nodes.
To illustrate the main idea, we assume $d_k/k$ to be an integer, and set $q_k = d_k/k$. In practice, the average degree $d_k$ can be a non-integer. To remedy this, we can introduce randomness to determine the value of $q_k$ for each new node such that the expectation of $q_k$ is $d_k/k$.

\begin{algorithm}[t]
\caption{Generating a random SF-like hypergraph}
\label{alg:sf-hypergraph}
\begin{algorithmic}[1]

\Require Number of nodes \(n\), highest order \(M\), average degrees \(\{d_k\}_{k=2}^{M}\)
\Ensure An unweighted hypergraph \(G=(V,E)\)

\State Choose an initial hypergraph \(G_0=(V_0,E_0)\) with \(n_0=|V_0|\geq M\)
\State Set \(G\leftarrow G_0\), \(V\leftarrow \{1,\ldots,n\}\)

\For{each \(k=2,\ldots,M\)}
    \State Initialize a multiset \(R_k\) by inserting each node \(u\in V_0\) \(\deg_k(u)\) times
\EndFor

\For{\(v=n_0+1,\ldots,n\)}
    \For{\(k=2,\ldots,M\)}
        \State Set \(q_k=d_k/k\)
        \State Draw \(Q_{k,v}=\lfloor q_k\rfloor+
        \operatorname{Bernoulli}(q_k-\lfloor q_k\rfloor)\)
        \State Set \(\mathrm{cnt}=0\)

        \While{\(\mathrm{cnt}<Q_{k,v}\)}
            \State Sample \(k-1\) distinct node labels
            \(T\subseteq\{1,\ldots,v-1\}\) from \(R_k\), with probability
            proportional to multiplicity
            \State Set \(e=T\cup\{v\}\)

            \If{\(e\notin E\)}
                \State Add \(e\) to \(E\)
                \State Insert every node in \(e\) once into \(R_k\)
                \State \(\mathrm{cnt}\leftarrow\mathrm{cnt}+1\)
            \EndIf
        \EndWhile

    \EndFor
\EndFor

\State \Return \(G=(V,E)\)

\end{algorithmic}
\end{algorithm}

Many social and biological networks, such as the citation network, are close to SF networks~\cite{barabasi1999emergence,barabasi2002evolution,gay2005innovation,hanaki2010dynamics}. The second row of~\cref{fig:ERSF} shows the degree distribution of SF-like hypergraph generated from our algorithm. The resulting degree distribution is approximately linear on a log-log plot for the parameter ranges used in our experiments, indicating heavy-tailed scale-free-like behavior.

\textbf{Regular-local hypergraph}: 
The ER-like and SF-like generators described above create global random hyperedges. In addition, we use a regular-local generator to create sparse bounded-locality instances. This generator is motivated by the fact that the cost of the conservative local certificates depends strongly on the size of the local boundary around a candidate group. A bounded-degree local construction provides a controlled setting in which this boundary size remains moderate as the number of nodes increases. Also, sparse Ising-like models with locally planted higher-order terms have recently been the subject of research in applications of QAOA algorithm on quantum hardware~\cite{Pelofske2024short, Pelofske2024scaling}.

The construction begins with a random \(d_{\mathrm{reg}}\)-regular graph on \(n\) nodes, generated using the standard random-regular-graph model. This graph is used as the pairwise backbone. Thus the number of pairwise edges is \(nd_{\mathrm{reg}}/2\), and the pairwise degree is exactly \(d_{\mathrm{reg}}\). The construction requires \(0\le d_{\mathrm{reg}}<n\) and \(nd_{\mathrm{reg}}\) even.

Higher-order interactions are then sampled locally. For each node \(v\), let \(B(v) := \{v\}\cup N(v)\) be the closed one-hop neighborhood of \(v\) in the regular backbone. Since the backbone is \(d_{\mathrm{reg}}\)-regular, \(|B(v)|=d_{\mathrm{reg}}+1\). Therefore an order-\(k\) local hyperedge can be sampled from such a neighborhood only if
 \(k \le d_{\mathrm{reg}}+1 \).
 
For each order \(k\ge 3\), the target number of local hyperedges is
\[
    N_k = \operatorname{round}\!\left(\frac{d_k n}{k}\right),
\]
so that the resulting average order-\(k\) degree is approximately \(d_k\), provided the requested density is feasible under duplicate rejection and local-degree constraints.

To avoid concentrating too many higher-order terms on a small number of nodes, the implementation also imposes a maximum higher-order degree. By default, this cap is set to
\[
    D_{\max}^{\mathrm{HO}}
    =
    \left\lceil \sum_{k=3}^{M} d_k \right\rceil + 1,
\]
and it counts the total number of higher-order hyperedges incident to a node, across all orders \(k\ge 3\). Candidate hyperedges that would violate this cap are rejected. If the requested density is too high, or if the local neighborhoods are too small, the generator may fail after a fixed retry budget.

\begin{algorithm}[t]
\caption{Generating a random regular-local hypergraph}
\label{alg:regular-local-hypergraph}
\begin{algorithmic}[1]

\Require Number of nodes \(n\), regular backbone degree \(d_{\mathrm{reg}}\), 
higher-order target degrees \(\{d_k\}_{k=3}^{M}\)
\Ensure An unweighted regular-local hypergraph \(G=(V,E)\)

\State Check that \(0\leq d_{\mathrm{reg}}<n\) and \(nd_{\mathrm{reg}}\) is even
\State Generate a random \(d_{\mathrm{reg}}\)-regular graph 
\(G_2=(V,E_2)\) on \(V=\{1,\ldots,n\}\)
\State Set \(E\leftarrow E_2\)

\For{each \(v\in V\)}
    \State Define the closed one-hop neighborhood
    \(B(v)\leftarrow\{v\}\cup N_{G_2}(v)\)
\EndFor

\If{there exists \(k\geq3\) with \(d_k>0\) and \(k>d_{\mathrm{reg}}+1\)}
    \State \textbf{stop}: an order-\(k\) edge cannot be sampled from a one-hop neighborhood
\EndIf

\State Set \(
D_{\max}^{\mathrm{HO}}
\leftarrow
\left\lceil \sum_{k=3}^{M} d_k \right\rceil+1
\)
unless another cap is specified

\State Set \(D^{\mathrm{HO}}(v)\leftarrow0\) for all \(v\in V\)

\For{\(k=3,\ldots,M\)}
    \If{\(d_k=0\)}
        \State \textbf{continue}
    \EndIf
    \State Set
    \(N_k\leftarrow\operatorname{round}\!\left(\frac{d_kn}{k}\right)\)
    \State Set \(\mathrm{cnt}=0\)
    \While{\(\mathrm{cnt}<N_k\)}
        \State Sample a center node \(c\in V\) uniformly at random
        \State Sample a \(k\)-element subset \(e\subseteq B(c)\) uniformly at random
        \If{\(e\notin E\) and
        \(D^{\mathrm{HO}}(u)<D_{\max}^{\mathrm{HO}}\) for all \(u\in e\)}
            \State Add \(e\) to \(E\)
            \For{each \(u\in e\)}
                \State \(D^{\mathrm{HO}}(u)\leftarrow D^{\mathrm{HO}}(u)+1\)
            \EndFor
            \State \(\mathrm{cnt}\leftarrow\mathrm{cnt}+1\)
        \EndIf
    \EndWhile

\EndFor

\State \Return \(G=(V,E)\)

\end{algorithmic}
\end{algorithm}

The purpose of regular-local generator is complementary to the ER-like and SF-like generators: it produces sparse higher-order instances with explicitly bounded local neighborhoods. Such instances are useful for testing the local-certification regime, because for fixed \(d_{\mathrm{reg}}\) and fixed maximum interaction order, candidate boundary sizes remain controlled as \(n\) grows.

We remark that there are various ways of generalizing ER and SF networks to higher orders. In some specific scenarios, other approaches may be more sensible. However, for our purpose of benchmarking the performance of GeneralHare algorithm, the current definitions allow us to draw a distinction between these two types of hypergraphs based on their features. Moreover, the ER/SF-like hypergraphs we propose here readily recover the normal ER/SF networks if we set $M=2$.

\subsection{Introduction to real-world data}
Here we give a brief introduction to the real-world datasets we use~\cite{benson2018}. 
Each dataset consists of a sequence of timestamped simplices. A simplex is a set of nodes, which can be interpreted as a hyperedge with unit weight. Each time a simplex reappears, we increase its weight by one. In this way, we convert each dataset into a hypergraph.
\begin{itemize}
	\item[-] Email networks (email-Enron; email-Eu): nodes are email addresses and a simplex is a set consisting of all recipient addresses on an email along with the sender’s address;
	email-Enron spans most of the duration of a company’s lifetime, and email-Eu spans over 2 years.
	\item[-] Drug networks from the National Drug Code Directory (NDC-classes):
	nodes are class labels (e.g., serotonin reuptake inhibitor)	and a simplex is the set of class labels applied to a drug (all applied at one time). \\
	\item[-] (NDC-substances): nodes are substances (e.g., testosterone) and a simplex is the set of substances in a drug; datasets include the complete history of the directory.
	\item[-] Drug usage in the Drug Abuse Warning Network (DAWN):
	nodes are drugs and a simplex is the set of drugs reportedly used by a patient prior to an emergency department visit.
	\item[-] Contact networks (contact-high-school; contact-primary-school): nodes are people and a simplex is a set of persons in close proximity to each other.
\end{itemize}

\section{Additional numerical experiments on reduction ratio and runtime}
\label{app:runtime}

\begin{figure}[tb]
	\centering
	\includegraphics[width=0.47\textwidth]{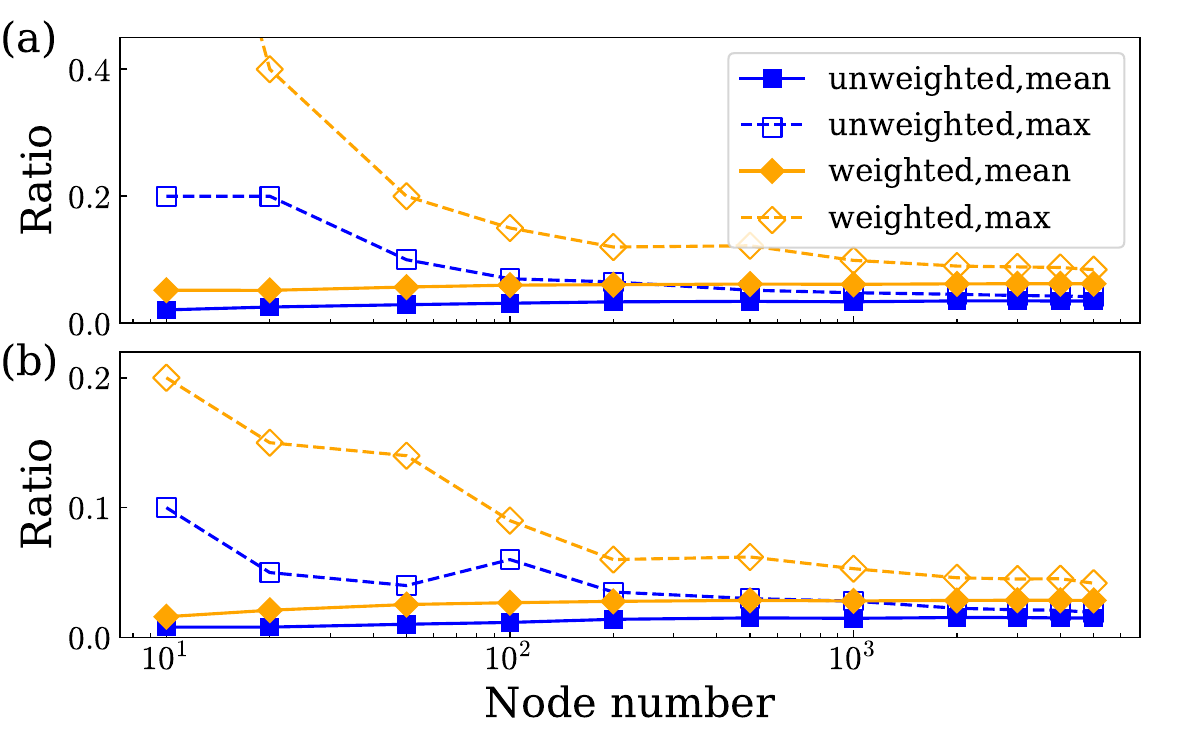}
	\includegraphics[width=0.47\textwidth]{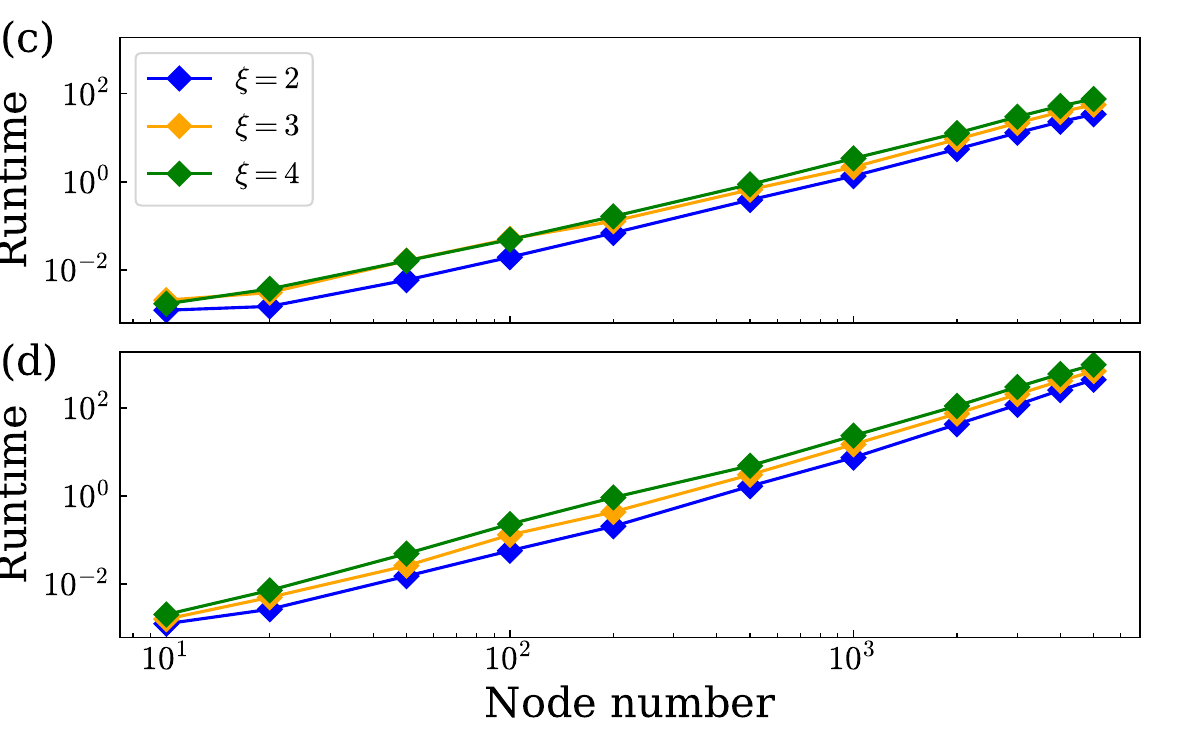}
	\caption{\textbf{Additional numerical experiments on reduction ratio and runtime.} \textbf{(a)} Reduction ratio with respect to node number on third-order weighted (orange line) and unweighted (blue line) ER-like hypergraphs, with $d_2=4,d_3=2$. For weighted hypergraph, $J$ is an integer chosen uniformly in $\{-4,-3,-2,-1,1,2,3,4\}$ for all hyperedges, while for unweighted hypergraph, $J=1$ for all hyperedges. The corresponding hollow markers denote the highest reduction ratio in 100 random samples for each case. When $n=10$, maximal ratio $1$ is out of scale. \textbf{(b)} Same as (a), but for fourth-order ER-like hypergraphs, $d_2=4,d_3=2,d_4=1$. \textbf{(c)} Processing time with different choices of $\xi$ on fourth-order weighted ER-like hypergraphs. \textbf{(d)} Same as (c), but for SF-like hypergraphs.}
	\label{fig:runtime}
\end{figure}

In~\cref{fig:runtime} (a) (b), we show the reduction ratio versus node number on our synthesized Erd{\H{o}}s-R{\'e}nyi hypergraphs. ER-like hypergraphs usually exhibit lower reduction ratios than SF-like hypergraphs of the same order and average degree.
In contrast to second-order case, the reduction ratio for small-sized ER-like hypergraphs ($n<100$) is surprisingly low. An intuitive explanation is that, with the existence of higher-order interactions, less information are left redundant.

To show the influence of user-chosen parameter $\xi$, we run GeneralHare on the same dataset for all valid choices of largest grouping size $\xi$, i.e. integers from 2 to the highest order $M$ of the Hamiltonian under consideration. We found that, although for extremely small-sized instances ($\approx 10$ nodes), larger $\xi$ results in higher reduction ratio, the advantage quickly vanishes as the node number increases. 

As for the processing time, we can infer from the linearity of log-log plot (\cref{fig:runtime} (c) (d)) that the time cost of GeneralHare is still polynomial with respect to number of nodes. Linear regression shows that the slope is approximately 1.75 for ER-like hypergraphs and 2 for SF-like hypergraphs.

In our synthesized dataset, for typical problem size (node number over 1000), $\xi$ larger than 2 shows no improvement in test examples, but induces an overhead in processing time. Thus we set $\xi=2$ in all our numerical experiments.
However, we still keep the criteria in general orders as they may be useful in other scenarios.

\section{GeneralHare-enhanced higher-order Ising  solver}
\label{sec:enhanced solver}

\subsection{Simulated bifurcation}
\label{app:SB}

Simulated bifurcation (SB) is an optimization algorithm that simulates adiabatic evolutions of classical nonlinear Hamiltonian systems exhibiting bifurcation phenomena \cite{goto2019combinatorial}. 
Because of its simultaneous updating, the acceleration of SB by massively parallel processing such as field-programmable gate array (FPGA) or graphics processing unit (GPU) is easier than for SA. 

In addition to the original adiabatic SB (aSB), two variants, named ballistic SB (bSB) and discrete SB (dSB), were proposed to enhance the power of SB in terms of solution accuracy~\cite{goto2021high}. We use bSB in this paper.

Let $x_i$ and $y_i$ denote, respectively, the position and momentum of a particle corresponding to the $i$-th spin. 
The equations of motion and effective Hamiltonian function for bSB are as follows~\cite{goto2021high}
\begin{align}
	\dot{x}_i & = \frac{\partial H_{\mathrm{bSB}}}{\partial y_i} = a_0 y_i 
	\label{bsb_x} \\
	\dot{y}_i & = -\frac{\partial H_{\mathrm{bSB}}}{\partial x_i} = -[a_0-a(t)] x_i -c_0 \left(h_i+\sum_{j\ne i}J_{ij}x_j \right) \label{bsb_y} \\
	H_{\mathrm{bSB}} & = \frac{a_0}{2} \sum_{i=1}^{n} y_i^2 + V_{\mathrm{bSB}} \\
	V_{\mathrm{bSB}} & = \frac{a_0-a(t)}{2} \sum_{i=1}^{n} x_i^2 + c_0 \left(\sum_{i=1}^{n} h_i x_i + \sum_{1\le i<j\le n} J_{ij}x_i x_j \right) \ \ \  \text{when } |x_i| \leq 1 \text{ for all } x_i \text{ , otherwise } V_{\mathrm{bSB}} = \infty,
	\label{eq:VbSB}
\end{align}
where $J_{ij}$ are the weights defined in the main text; $a_0$ and $c_0$ are positive constants, and $a(t)$ is a control parameter increased from zero.

This approach can be extended to higher-order polynomial cost functions by replacing the quadratic Ising energy with its higher-order polynomial analogue. In this case, Eqs.~(\ref{bsb_x}-\ref{eq:VbSB}) will read as follows

\begin{align}
	&\dot{x}_i  = \frac{\partial H_{\mathrm{bSB}}}{\partial y_i} = a_0 y_i ,
	\label{eq:bsb_x_gen} \\
	&\dot{y}_i  = -\frac{\partial H_{\mathrm{bSB}}}{\partial x_i}
    =
    -\bigl[a_0-a(t)\bigr]x_i -
    c_0  \sum_{\substack{I\in E\\ i\in I}}  J_I  \prod_{j\in I\setminus\{i\}} x_j , 	\label{eq:bsb_y_gen} \\
	&H_{\mathrm{bSB}}(\mathbf{x},\mathbf{y},t)
     =
    \frac{a_0}{2}\sum_{i=1}^n y_i^2 + V_{\mathrm{bSB}}(\mathbf{x},t), \\
	&V_{\mathrm{bSB}}(\mathbf{x},t)
     =
    \begin{cases}
    \displaystyle
    \frac{a_0-a(t)}{2}\sum_{i=1}^n x_i^2
    +
    c_0 \sum_{I\in E}
    J_I \prod_{j\in I} x_j,
    &
    \text{if  } |x_i|\le 1 \ \text{for all } i, \\[1.0em]
    +\infty,
    &
    \text{otherwise}.
    \end{cases}
    \label{eq:VbSB_gen}
\end{align}

In \cref{eq:bsb_y_gen} we take into account the assumption that Hamiltonian does not have self-interaction, as it was discussed in the `Preliminaries' section in the main text. We perform differentiation by simply dropping all terms that do not contain $x_i$, and dropping $x_i$ from those where it is present.

This formulation of SB retains its advantage in that all variables are updated, based only on values from the previous iteration. So, when properly implemented, it still allows to update all variable values simultaneously, although the force term in Eqs.~(\ref{bsb_x}-\ref{bsb_y}) is no longer linear in \(\mathbf{x}\).

\begin{figure}
    \centering
    \includegraphics[width=0.9\textwidth]{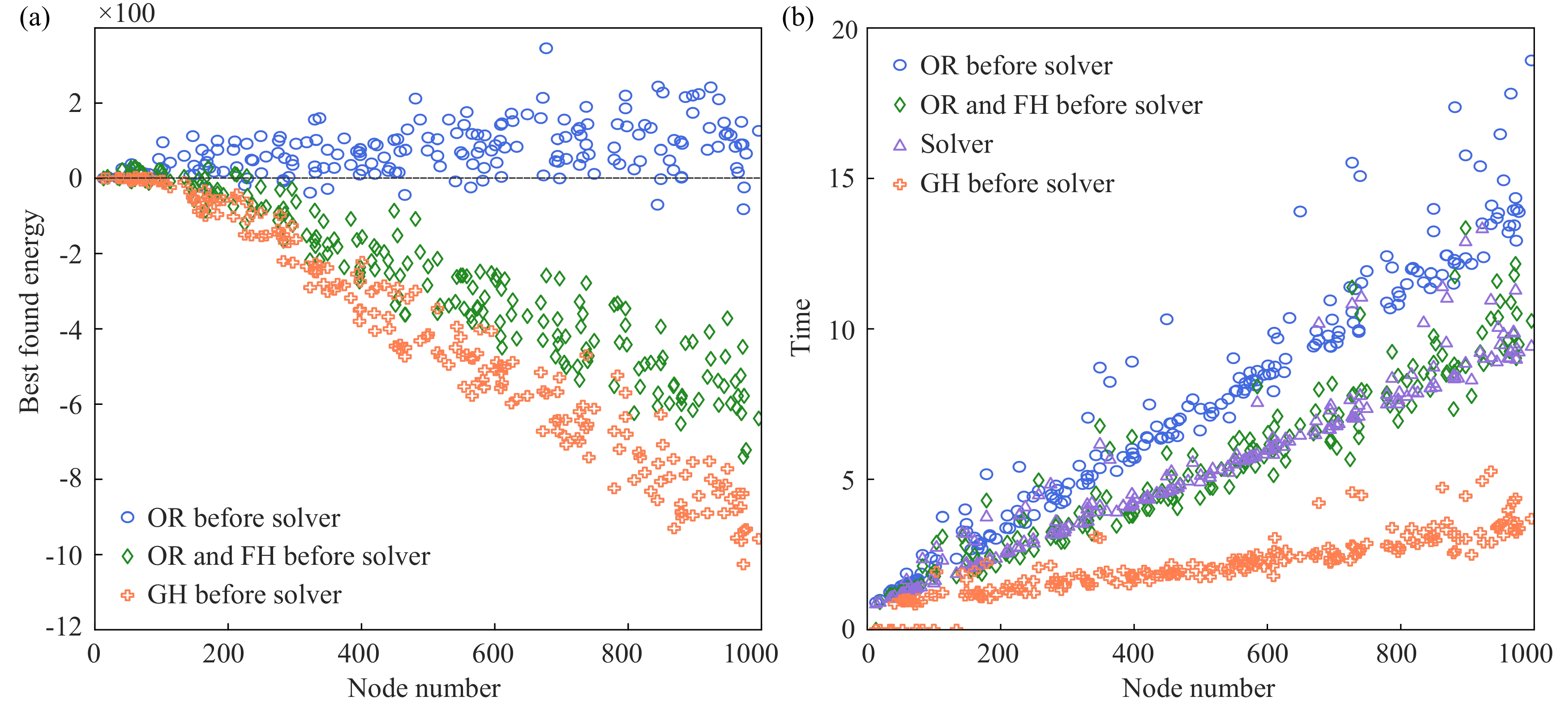}
    \caption{\textbf{Performance comparison of the GeneralHare-enhanced solver with solvers equipped with other preprocessing techniques.} We randomly generated 200 third-order SF-like hypergraphs of sizes ranging from $10$ to $1000$ nodes. Each data point denotes a sample. The average degrees are set to be $d_2=2,d_3=0.5$. \textbf{(a)} Best energy found by the solver versus node number. (We have set the results obtained by the standard solver as baseline.) \textbf{(b)} Runtime versus node number. `FH' means FastHare, and `GH' means GeneralHare.}
    \label{fig:solver_sb}
    \includegraphics[width=0.9\textwidth]{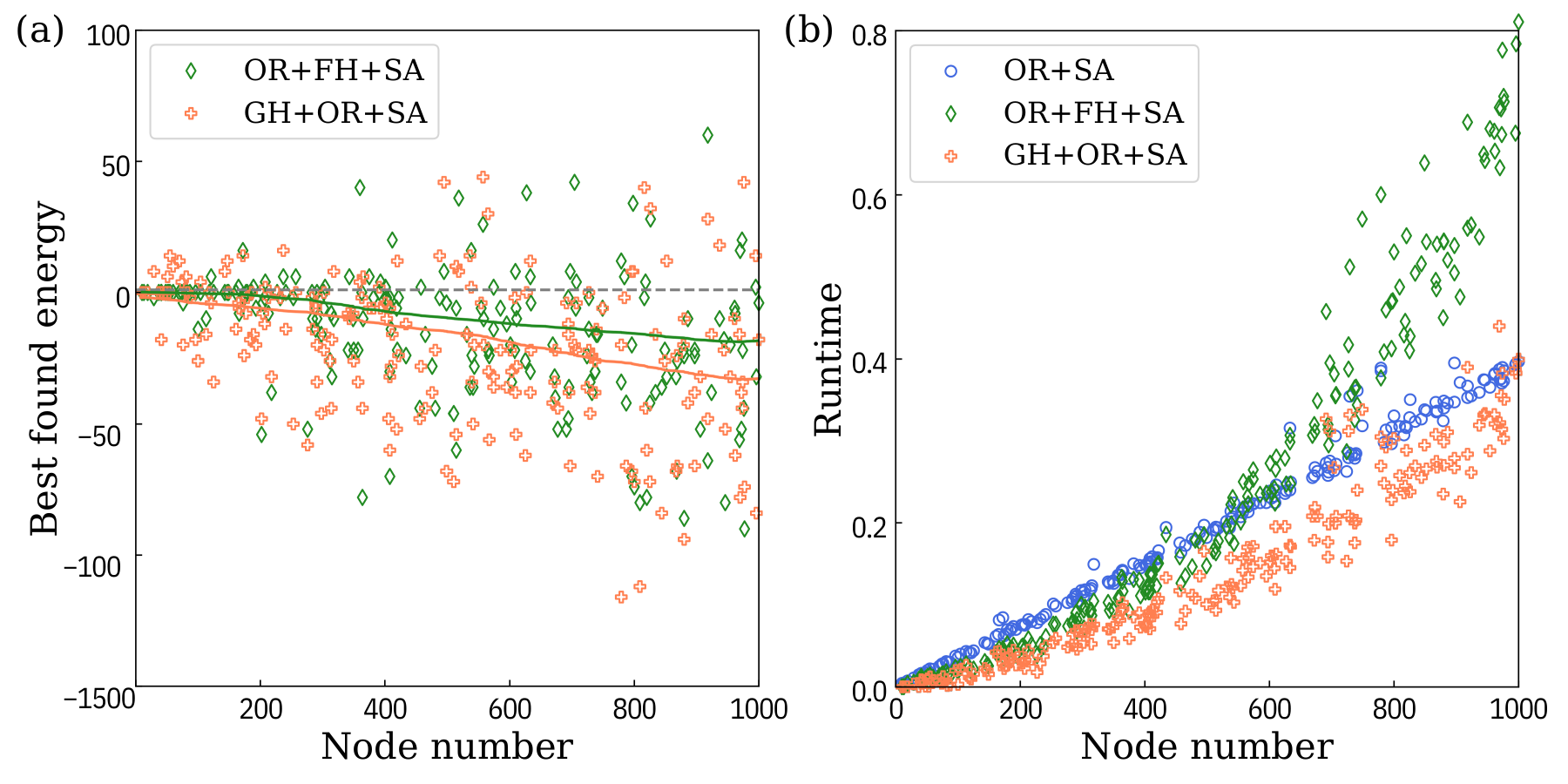}
	\caption{\textbf{Performance comparison of GeneralHare-enhanced solver with solvers equipped with other preprocessing techniques.} We randomly generated 200 third-order SF-like hypergraphs of sizes ranging from $10$ to $1000$ nodes. Each data point denotes a sample. The average degrees are set to be $d_2=2,d_3=0.5$. \textbf{(a)} Best energy found by the solver versus node number. (We have set the results obtained by the standard solver as baseline.) The orange and green lines denote the moving average of 500 random samples (200 out of which are shown as data points in the figure) solved by the corresponding methods. \textbf{(b)} Runtime versus node number. We set the number of sweeps in SA as 1000, and the number of reads as 10.}
	\label{fig:solver_sa}
\end{figure}

\subsection{GeneralHare-enhanced higher-order Ising  solver}
\label{app:enhanced_solver}
Heuristic Ising solvers provide no guarantee of finding the exact ground state of a given Hamiltonian. Preprocessing techniques such as Hamiltonian reduction would simplify the Hamiltonian, and improve the final solution. This can be demonstrated by comparing the quality of solutions obtained from the standard solver and the GeneralHare-enhanced solver.

Here we take generalized ballistic simulated bifurcation (bSB)~\cite{goto2019combinatorial, goto2021high} and Simulated Annealing (SA)~\cite{kirkpatrick1983optimization} as standard solvers. Both are widely used to search for low-energy states of Ising models. For comparison, we also adopt order reduction (OR) in our experiment with the PyQUBO module~\cite{Zaman2022pyqubo}. We evaluate the quality of the solution $\mathbf{\sigma}$ by the corresponding energy $H(\mathbf{\sigma})$. The lower the energy, the better the solution. We also record the end-to-end runtime.

To comprehensively demonstrate the advantage of our scheme, on \cref{fig:solver_sb} we obtain the solution for each instance in the test dataset with the following four different approaches: (1) solve directly with bSB; (2) apply GeneralHare on the instance and solve the reduced Hamiltonian with bSB; (3) first apply order reduction, then solve with bSB; (4) first apply order reduction, then FastHare, and solve the resulting Hamiltonian with bSB. 

 We benchmark the four approaches described above with third-order SF-like hypergraphs of size ranging from 10 to 1000 nodes.   In \cref{fig:solver_sb}~(a), we set the energy obtained by the standard solver as baseline, and record the difference between energy obtained by other approaches and the baseline. For all pipelines involving reduction, the plotted energy is the value of the original Hamiltonian evaluated on the reconstructed original spin configuration, not the raw objective value of the reduced Hamiltonian. In \cref{fig:solver_sb}~(b) we show the runtime for four approaches, among which the GeneralHare-enhanced solver is the fastest. Compared to the direct application of higher-order SB solver, order reduction actually makes the solution quality worse. On the other hand, both FastHare and GeneralHare can enhance the performance, and the GeneralHare-enhanced solver can obtain better solutions with even shorter runtime. Moreover, the larger the problem size (node number), the more evident the advantage of the GeneralHare-enhanced solver.

To eliminate the influence of the choice of standard solver on benchmarking, in~\cref{fig:solver_sa} we take order reduction (OR) plus simulated annealing (SA) implemented by D-Wave as the standard solver, and compare different preprocessing techniques using the same dataset as~\cref{fig:solver_sb}. The result again shows that GeneralHare-enhanced solver outperforms the others.

From the reference lines in~\cref{fig:solver_sa}~(a) we see that in terms of solution quality, GeneralHare-enhanced solver outperforms OR+SA as well as the  FastHare-enhanced one on average. From~\cref{fig:solver_sa}~(b), we find that the runtime of GeneralHare-enhanced solver, being the lowest of the three, increases almost linearly with respect to node number, while the runtime of FastHare-enhanced solver shows superlinear growth as the problem size increases. 

The dataset we choose consists of rather sparse hypergraphs with unevenly distributed node degrees, thus allowing Hamiltonian reduction techniques (GeneralHare and FastHare) to take effect. For this specific dataset, we see that GeneralHare-enhanced solver outperforms its counterparts equipped with other preprocessing techniques, regardless of the standard solver we choose. For denser graphs or those with more evenly distributed node degrees, the advantage may decline.

\section{Influence of candidate-size cutoff parameter $\xi$ on reduction ratio and runtime}
\label{app:xi_sweep}

The parameter \(\xi\) controls the maximum size of candidate groups considered by the reduction procedure. In our implementation, candidates are generated only from hyperedges \(X\) satisfying \(|X|\leq \xi\). This restriction is used for computational efficiency: increasing \(\xi\) allows the algorithm to test larger candidate groups, but also increases the cost of local certification. The resulting procedure should therefore be interpreted as safe but incomplete with respect to this candidate set: increasing \(\xi\) can reveal additional reducible groups, whereas a smaller \(\xi\) may miss them, but certified reductions found within the searched candidate set remain valid.

To examine this trade-off, we performed an additional \(\xi\)-sweep on regular-local higher-order hypergraphs using the demonstrator implementation GH\_minimal. This instance family was chosen because it provides a controlled bounded-locality setting while still containing genuine higher-order interactions. We considered two profiles: rather sparse higher-order interactions over a denser second-order graph layer, and denser third-order interaction layer over sparse pairwise graph. In detail, the first profile has maximum order \(M=4\), with a random \(d_{\mathrm{reg}}=3\) regular pairwise backbone and target local higher-order degrees \(d_3=0.5\), \(d_4=0.3\). The second profile has maximum order \(M=3\), with a sparser \(d_{\mathrm{reg}}=2\) regular pairwise backbone and denser third-order component \(d_3=2\). For each profile, we generated \(10\) independent random instances using different seeds and applied GH\_minimal with \(\xi=2,3,4\) for the \(M=4\) profile and \(\xi=2,3\) for the \(M=3\) profile. The number of nodes was \(n=120\), and all hyperedge weights were sampled independently from \(\{-4,-3,-2,-1,1,2,3,4\}\).

\begin{figure}[t]
    \centering
    \includegraphics[width=1.0\linewidth]{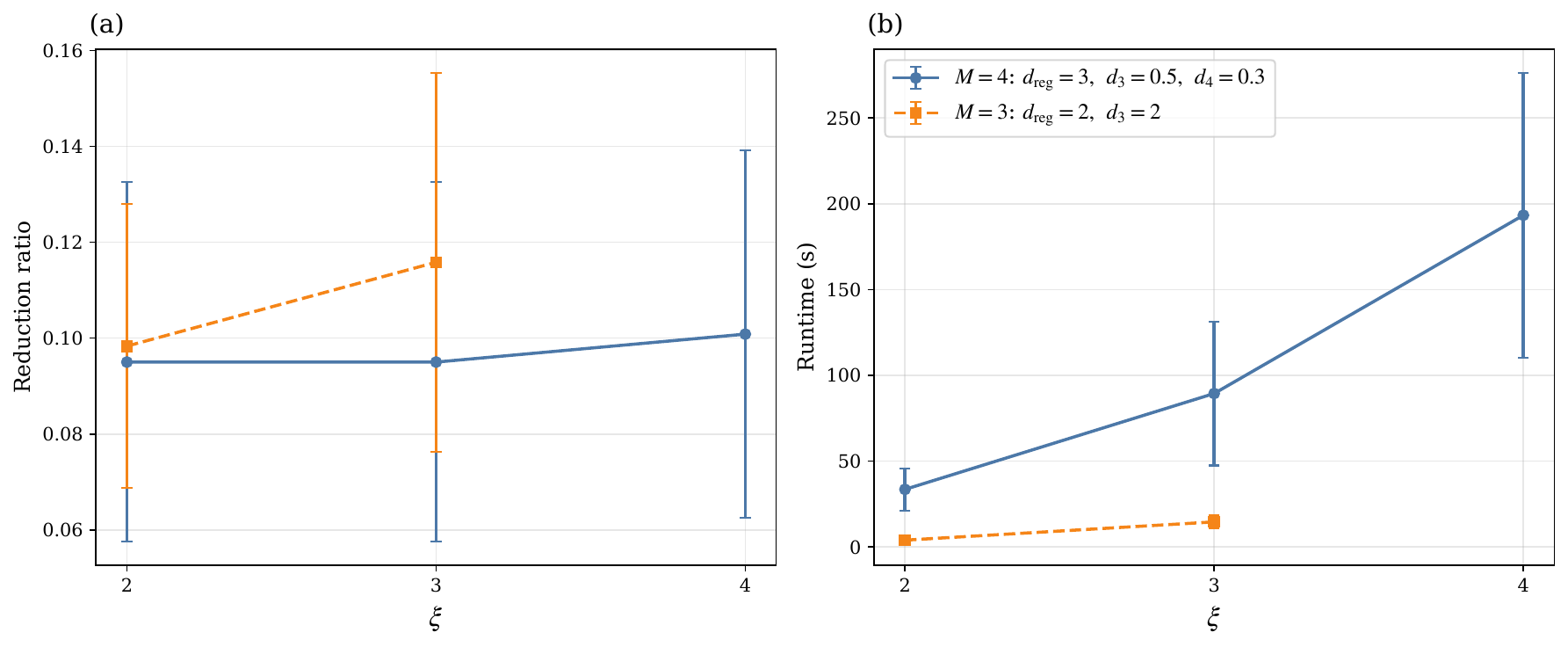}
    \caption{
    Influence of the candidate-size cutoff \(\xi\) on GH\_minimal for regular-local higher-order hypergraphs.
    (a) Reduction ratio and (b) runtime as functions of \(\xi\).
    Markers show averages over \(10\) independent random instances and error bars show one standard deviation.
    Two regular-local profiles are shown: four-order hypergraphs with \(d_{\mathrm{reg}}=3\), \(d_3=0.5\), \(d_4=0.3\), and three-order instances with \(d_{\mathrm{reg}}=2\), \(d_3=2\).
    }
    \label{fig:xi_sweep_regular_local}
\end{figure}

The results are shown in Fig.~\ref{fig:xi_sweep_regular_local}. In both profiles, increasing \(\xi\) gives only a modest improvement in the average reduction ratio, while the runtime increases substantially, especially for the \(M=4\) profile. This supports the choice \(\xi=2\) as a practical default in the experiments: it captures most of the observed reduction at much lower computational cost. The \(M=3\) profile at the same time suggests that selecting a higher value of $\xi$ may be justified for problems dominated by higher-order interactions.

\section{Example of compression and reconstruction}
\label{app:compression_reconstruction_example}

We illustrate the compression rule and reconstruction map using the toy
Hamiltonian shown in Fig.~\ref{fig:GEJ}. Consider
\begin{equation}
\begin{aligned}
H(\boldsymbol{\sigma})={}&
J_a\sigma_a+J_b\sigma_b+J_c\sigma_c+J_d\sigma_d
+J_{ab}\sigma_a\sigma_b
+J_{bc}\sigma_b\sigma_c
+J_{ac}\sigma_a\sigma_c  \\
&+J_{abc}\sigma_a\sigma_b\sigma_c
+J_{cd}\sigma_c\sigma_d .
\end{aligned}
\end{equation}
Suppose that the group \(X=\{a,b,c\}\) is compressed with relative
configuration
\[
    \boldsymbol{\eta}_X=(+1,+1,+1).
\]
Introducing the merged spin \(\tau_x\), the reconstruction map for this
compression is
\[
    \sigma_a=\tau_x,\qquad
    \sigma_b=\tau_x,\qquad
    \sigma_c=\tau_x,
\]
while the external spin remains unchanged, \(\sigma_d=\tau_d\). Substitution
gives
\begin{equation}
\begin{aligned}
H(R(\boldsymbol{\tau}))={}&
(J_a+J_b+J_c+J_{abc})\tau_x
+J_{cd}\tau_x\tau_d
+J_d\tau_d  \\
&+\left(J_{ab}+J_{bc}+J_{ac}\right).
\end{aligned}
\end{equation}
Thus the third-order term becomes a linear term in the merged spin, the
external pairwise term \(J_{cd}\sigma_c\sigma_d\) becomes a pairwise term
\(J_{cd}\tau_x\tau_d\), and the internal pairwise terms become the additive
constant
\[
    C=J_{ab}+J_{bc}+J_{ac}.
\]
This constant may be stored explicitly or omitted during optimization, since it
does not affect the minimizing configurations. If it is stored, then
\[
    H(R(\boldsymbol{\tau}))=H_{\mathrm{red}}(\boldsymbol{\tau})+C
\]
for every reduced configuration \(\boldsymbol{\tau}\). If it is omitted, the
energy of a candidate solution can be computed by reconstructing the full
spin configuration and evaluating the original Hamiltonian \(H\).

In the released GH\_minimal implementation, reconstruction is tracked
conceptually by a node map and a sign map. Let \(\pi\) denote the map from
original nodes to current reduced nodes, and let \(s_i\in\{-1,+1\}\) denote the
relative sign of original spin \(i\) with respect to its current reduced spin.
For an unfixed original node \(i\), reconstruction has the form
\[
    \sigma_i=s_i\,\tau_{\pi(i)} .
\]
For the compression above, the tracked maps are
\[
    \pi(a)=\pi(b)=\pi(c)=x,\qquad
    s_a=s_b=s_c=+1 .
\]
Therefore, if the solution of the reduced problem has \(\tau_x=-1\), then the reconstructed
values of the compressed group are
\[
    \sigma_a=\sigma_b=\sigma_c=-1 .
\]

Node fixation is represented in the same reconstruction logic. If, after the
compression, node \(d\) is certified and fixed to \(\sigma_d=-1\), we may write
\[
    \pi(d)=\bot,\qquad s_d=-1,
\]
where \(\bot\) indicates that the original node no longer corresponds to a
free reduced spin and \(s_d\) stores its fixed value. The full reconstruction
from a reduced assignment \(\tau_x\) is then
\[
    \sigma_a=\tau_x,\qquad
    \sigma_b=\tau_x,\qquad
    \sigma_c=\tau_x,\qquad
    \sigma_d=-1 .
\]
For example, if the final reduced solution is \(\tau_x=+1\), the reconstructed
original configuration is
\[
    (\sigma_a,\sigma_b,\sigma_c,\sigma_d)=(+1,+1,+1,-1).
\]
If the final reduced solution is \(\tau_x=-1\), the reconstructed original
configuration is
\[
    (\sigma_a,\sigma_b,\sigma_c,\sigma_d)=(-1,-1,-1,-1).
\]
Thus the reduced problem is used only to determine the remaining free reduced
spins, while the tracked node map, sign map, and fixed-node information define
the corresponding configuration of the original Hamiltonian.

\end{document}